\documentclass[usenatbib]{mnras}
\usepackage{epsfig}
\usepackage{times}
\usepackage{amsmath,amssymb}
\usepackage{xcolor}
\usepackage{tikz}
\graphicspath{{figures/}}
\usepackage{cprotect}

  \newcommand{\Teff}{\mbox{\,\em T$_{\rm eff}$}}         

  \newcommand{\kmsec}{\,\mbox{$\mbox{km}\,\mbox{s}^{-1}$}}    
  \def\simge{\mathrel{\raise1.16pt\hbox{$>$}\kern-7.0pt
    \lower3.06pt\hbox{{$\scriptstyle \sim$}}}}           
  \def\simle{\mathrel{\raise1.16pt\hbox{$<$}\kern-7.0pt
    \lower3.06pt\hbox{{$\scriptstyle \sim$}}}}           

\newcommand{\appropto}{\mathrel{\vcenter{
  \offinterlineskip\halign{\hfil$##$\cr
    \propto\cr\noalign{\kern2pt}\sim\cr\noalign{\kern-2pt}}}}}

\usepackage{tabu}   
\newcolumntype{L}[1]{>{\raggedright\let\newline\\\arraybackslash\hspace{0pt}}m{#1}}
\usepackage{afterpage}
\usepackage{caption} 
\usepackage{pdflscape} 
\usepackage{rotating} 
\DeclareGraphicsExtensions{.pdf,.png,.jpg,.eps,.ps}
\usepackage{breqn}   
\usepackage{soul}  
\usepackage{stfloats}
\usepackage[linesnumbered,ruled,lined]{algorithm2e}  

\title[SALT hot subdwarfs: PCA \& kinematics]{The SALT survey of helium-rich hot subdwarfs: unsupervised classification and kinematic analysis}

\author[A.~{Philip Monai} \& C.~S.~Jeffery]{{A.~{Philip Monai}$^{1,2}$\thanks{email: asish.philip.monai@armagh.ac.uk}} and {C.~S.~Jeffery$^{2}$}
\\
$^{1}$School of Mathematics and Physics, Queen's University Belfast, Belfast, BT7 1NN, UK\\
$^{2}$Armagh Observatory and Planetarium, College Hill, Armagh, BT61 9DG, UK
}
\date{Last updated ; in original form}

\pubyear{2025}

\begin{document}
\label{firstpage}
\pagerange{\pageref{firstpage}--\pageref{lastpage}}
\maketitle

\begin{abstract}
Hydrogen-deficient stars form sequences across the HR diagram from cool R CrB stars to helium-rich hot subdwarfs and PG1159 stars, tracing multiple evolution pathways. 
The Southern African Large Telescope (SALT) has been used to conduct a medium-resolution spectroscopic survey of 697 helium-rich hot subdwarfs and related stars. 
Using 587 stars from the full SALT sample, we define an unsupervised data-led classification system based on Principal Components Analysis (PCA) and Spectral Clustering (SC), designed as a data-discovery tool to identify possible new groups and sequences within the data. 
Using the first 3 principal components (PCs), we identify 6 major clusters of hot subdwarf spectra, identified with the traditional spectral classes of classical sdB and sdO stars, helium-rich He-sdO stars and very hot subdwarfs. 
Two clusters covering large volumes of PC space include a) He-sdB and He-sdOB stars and b) intermediate iHe-sdB and iHe-sdOB stars. 
Most spectra in these major clusters form connected sequences in PC space. 
Using a second stage of PCA and SC, we identify sub-clusters within 4 major clusters, particularly in the He-sdB/OB and iHe-sdB/OB clusters. 
In comparison with traditional MK-like classification, we find that the PC clusters are reasonably well separated, with some overlap at cluster boundaries. 
For very hot sdOs, where the number of standards is small, PC classification has led to a revision of the Drilling MK-like system. 
{\it Gaia} DR3 measurements have been used to determine kinematic parameters for the PC-defined classes. 
Although disk stars account for a significant fraction of all classes, He-sdOs and Hot-sdOs include a higher fraction of halo stars.

\end{abstract}



\section{Introduction}
\label{sect:intro}
Hot subluminous stars are stars of spectral types O and B that lie below the upper main sequence of the Hertzsprung-Russell diagram, occupying the extreme blue end of the blue horizontal branch \citep{heber16}. 
They were first discovered through photometric surveys of faint blue stars \citep{humason47} and their numbers substantially increased by the Palomar-Green survey of faint blue objects \citep{green86}. 
Spectroscopically they may be broadly identified as hot subdwarf B (sdB), subdwarf OB (sdOB) and subdwarf O (sdO) stars \citep{drilling03}. 
From a volume-complete sample of 305 hot subdwarfs, \citet{dawson24} find population fractions of $\sim58\%$ sdBs, $\sim21\%$ sdOBs, $\sim11\%$ sdOs and $\sim10\%$ helium rich hot subdwarfs. 
The latter is comparable with the  helium-rich fraction $\sim\,10\%$ found by \citet{geier17}.  
A significant proportion of the helium-rich hot subdwarfs are also carbon and/or nitrogen enriched \citep{stroeer07,naslim10}. 

A number of formation channels have been proposed for hot subdwarfs from theoretical modelling and binary star population synthesis. 
The large sample of close binary hot subdwarfs with periods less than 10 days \citep{geier22,schaffenroth22,schaffenroth23} provides evidence of common-envelope ejection from a red-giant with a less massive white dwarf or M-dwarf companion shortly prior to core helium ignition \citep{han02,han03}. 
Hot subdwarfs in wide binaries with more massive G or K companions \citep{jeffery98b,vos19} indicate a stable Roche Lobe overflow channel \citep{han02,han03}.
\citet{webbink84} and \citet{iben90} suggested that the merger of two helium white dwarfs could produce an isolated hot subdwarf; \citet{saio00} and \citet{zhang12a} argued that these would have helium-rich surfaces. 

The first catalogue of 1225 spectroscopically identified hot subdwarfs was compiled by \citet{kilkenny88}. 
Since the advent of large-scale sky surveys such as the Sloan Digital Sky Survey \citep{kepler16,kepler19,kepler21}, {\it Gaia} \citep{geier17,geier20,culpan22} and LAMOST \citep{luo16,luo19,luo21,luo24}, the discovery of hot subdwarfs has continued to increase. 

Several spectral classification systems have been used for low-resolution spectra ($R\sim10$\AA) of hot subdwarfs; early systems were reviewed by \citet{drilling13} (D13 hereafter). 
Following recent convention \citep{geier17,culpan22}, we here use the \citet{moehler90a} classes of sdB, sdOB, sdO, He-sdB and He-sdO, to which we add the intermediate designations iHe-sdB and iHe-sdOB. 
Classes sdB, sdOB and sdO refer to early type stars with strong Balmer lines intermediate between main-sequence stars and white dwarfs, with sdB having weak He{\sc i}, sdOB showing He{\sc i} {\it and} He{\sc ii} and  sdO showing weak He{\sc ii} lines. In the absence of He{\sc ii} the boundary between sdO and sdB is based on Balmer lines alone. 
Hot subdwarf classes indicating very weak Balmer lines are prefixed `He-'; thus He-sdB, He-sdOB and He-sdO. 
\citet{moehler90a} defines He-sdB having strong He{\sc i} accompanied by weak He{\sc ii} and He-sdO having strong He{\sc ii} along with weak He{\sc i}.
Here the r\^oles of He{\sc i} and He{\sc ii} are crucial, but definitions in the literature are not well established; {\it e.g.} \citet{lei19} define He-sdOB as having dominant He{\sc i} lines along with He{\sc ii} and Balmer lines.
For this paper we will consider boundaries between He-sdB and  He-sdOB and between He-sdOB and He-sdO to be indicated approximately by He{\sc ii} 4686/He{\sc i} 4713 $\approx 1$  and  He{\sc ii} 4686/He{\sc i} 4471 $\approx 1$ respectively.
At $\sim3\%$ of all hot subdwarfs\footnote{8 out of 305 hot subdwarfs listed by \citet{dawson24} would be classified iHe-sd.}, the intermediate helium-rich hot subdwarfs (iHe-sdOB) show significant Balmer, He{\sc i} and, usually, He{\sc ii} lines. 
The designation iHe-sdB may be used if He{\sc ii} is weak\footnote{On the basis of He/H ratios derived from spectral analyses, \citet{luo21} adopted He-poor (pHe), He-weak (wHe), intermediate helium (iHe) and extreme helium (eHe) designations.}.

For intermediate dispersion spectra ($R\sim2$\AA) D13 developed a three-dimensional MK-like system which adds a `helium class' to the conventional MK spectral and luminosity classes \citep{morgan43,keenan87,morgan84}.  
Carbon and nitrogen-rich spectra are also indicated. 
Like all MK classification schemes, D13 use standard stars to define spectral classes; classification of additional stars is achieved by  comparison against these standards.  
In practice, one uses the ratios of the depths and widths of specific absorption lines obtained at a given spectral resolution and wavelength range. \citet{jeffery21a} used D13 to classify 107 He-rich hot subdwarfs observed with the Southern African Large Telescope (SALT), while \citet{zou24} have provided D13-based classifications for 1224 hot subdwarfs from LAMOST.  
Such classification schemes give a precise description of the spectrum and, if calibrated against model atmosphere analyses, can provide estimates for surface temperature (spectral class),  gravity (luminosity class) and helium-to-hydrogen ratio (helium class). 

Given that spectral classifications depend on largely historical criteria from relatively small samples, the question arises whether any single method can accurately capture {\it a priori} all of the information available in large new surveys and whether they can usefully identify groups or sequences of stars with connected properties. 
For example, both D13 and \citet{jeffery21a} demonstrate that hot subdwarfs occupy a characteristic locus in a plot of helium versus spectral class; but it is more difficult to discern the r\^ole of the luminosity class.  
A critical shortcoming of the D13 system is a shortage of standards for very early-type spectra, especially around sdO1 -- sdO5 and, for He-rich stars, for spectral types later than sdB1 \citep{drilling13}.  
Another approach is to carry out model atmosphere analyses and identify patterns in temperature, gravity and composition, but this depends on the extent and validity of the model atmosphere space in addition to the quality of the spectra, and has a different overall goal \citep{jeffery21a,dorsch25}. 

A third approach is to use machine-learning techniques to establish whether other common features or patterns are present.
One attempt to automate the classification of hot subdwarfs using spectra from the D13 sample and from the SDSS used the method of principal components analysis (PCA) and artificial neural networks \citep{winter04,winter06}. 
More recent applications use convolutional neural networks \citep{tan22}, residual neural network, support vector machine \citep{cheng24} and self-organizing maps \citep{vazquez24} or any combination of these methods as tools for data discovery applied to {\it Gaia} and LAMOST data. 
All of these have been carried out using samples dominated by hydrogen-rich hot subdwarfs. 
In practice, spectral diversity is much greater among hydrogen-deficient hot subdwarfs \citep{jeffery21a}. 

The goal of the present paper is twofold: (1) To apply machine-learning techniques to a new sample of hot subdwarf spectra, in which hydrogen-weak or hydrogen-deficient spectra are strongly represented, to generate a  data-led classification system applicable to all single- spectrum hot subdwarfs, and to use it as a discovery tool within the present survey and in future work. (2) To perform kinematic analysis of the hot subdwarf sample using {\it Gaia} positional and proper motion data.

\section{Data Preparation}
\label{sect:SALT}
The primary data for this study are spectra of 697 hot subdwarfs and related objects obtained with the Robert Stobie Spectrograph (RSS: resolution $R \sim 3\,600$, \citet{burgh03, kobulnicky03}) on the Southern African Large Telescope (SALT). 
The sample is an extension of the 100-star sample presented by \citet{jeffery21a} and the data are described in full by \citet{jeffery25b}.

For analysis in this paper, the spectra are normalised, shifted to rest wavelength, and resampled to a grid of 4000 flux bins equally spaced in wavelength between 3850 and 5100 \AA. Radial velocities are measured by cross-correlation with appropriate theoretical templates using the cross-correlation function and template matching algorithms from {\sc SPECUTILS} \citep{astropy22}. 

To simplify automated classification, it is necessary that sample spectra be uncontaminated by, for example, a nearby cool star. 
Otherwise the classification system would require additional dimensions for which insufficient training data are available. 
68 stars were excluded from the full sample of 697 for the following reasons:
8 stars were identified by \citet{jeffery25b} as having strong stellar Ca H and K absorption and/or the 4430 G band. 
A few stars with high S/N spectra show a weak cool star spectrum, {\it e.g.} = J225635.9-524836 = EC 22536-5304 \citep[][Star \#673]{jeffery25b}. 
The latter is a well-known long-period binary \citep{dorsch21}. 
Including these stars, 37 were identified as having an red/infrared excess from their spectral energy distributions, and therefore either do or are likely to have composite spectra: {\it e.g.} J221656.0-643150 = BPS CS 22956-0094 \citep{snowdon25}.
Further details will be given by Dorsch et al. (in preparation).
18 stars lie outside the scope of the sample, including white dwarfs, PG1159 stars, cataclysmic variables, [WELS] and RCrB stars. 
5 were excluded because their spectra are incomplete. 
A list of the excluded stars is given in Table\,\ref{tab:excluded}.

\section{Methods}

Dimensionality reduction is the process of reducing the number of features in a dataset by either selecting the features that best describe the dataset or by constructing a new set of features to represent the original dataset. The high dimensionality of stellar spectra is a problem that has been addressed to a significant degree in recent years. Several combinations of dimensionality reduction and clustering algorithms were considered. For dimensionality reduction, these included principal components analysis (PCA), Uniform Manifold Approximation and Projection (UMAP) and some autoencoders. PCA was preferred because it provides a linear and interpretable transformation, enabling reconstruction of spectra and rejection of noisy data. 
In contrast, UMAP does not offer a straightforward inverse mapping for reconstruction and autoencoders require large, balanced training sets and are impractical for the present sample. 
For clustering, we tested k-means, Agglomerative Clustering, Spectral Clustering, Gaussian Mixture Models (GMM) and Variational Bayesian GMM. 
After several trials, Spectral Clustering proved most effective for our non-uniform sample, as it does not assume spherical clusters and adapts well to complex distributions. 
\begin{table*}
\caption[]{The 152 stars with good $S/N$ ratio in the standard sample given by their SALT designation (second column), corresponding to position (J2000.0). The first column indicates star \# from \citet[][table A.1]{jeffery25b}. Their corresponding D13 spectral type are also provided from \citet[][:\ddag]{drilling13}, \citet[][:\dag]{jeffery21a} and \citep{jeffery25b}. The 12 EHes in the standard sample have been marked with *.}
\label{t:training}
    \begin{tabular}{c L{4cm} L{4cm} c L{4cm} l }
\hline
\# & SALT & Drilling class & \# & SALT & Drilling class\\
\hline
3 & J000431.0-242621 & sdB0.5VII:He40 & 256 & J125657.1-601950 & sdBN0IV:He31\\
6 & J001853.2-315601\,* & sdBC0.2VI:He37\,\dag & 268 & J131748.1-284802 & sdB0.2VII:He37\\
16 & J005509.4-722756 & sdB2.5II:He15 & 271 & J132259.6-383813 & sdO2VII:He03\\
19 & J011338.2-151102 & sdBC1IV:He31 & 276 & J133146.3-194825 & sdOC9.5VII:He39\,\dag\\
20 & J011652.9-221208 & sdB2.5II:He24\,\dag & 281 & J134648.7-402538 & sdON7.5VI:He35\\
21 & J011919.7-581808 & sdO8VII:He01 & 283 & J134902.3-352438 & sdO6VII:He10\\
31 & J014307.5-383316 & sdOC7.5VII:He40\,\dag & 293 & J135957.2-630009 & sdO8.5V:He28\\
42 & J021834.5-641050 & sdB0.2VI:He38 & 296 & J140342.9-543031 & sdO4VI:He02\\
46 & J023326.1-591231 & sdOC5VI:He38\,\dag & 301 & J141159.3-362740 & sdO4VI:He02\\
48 & J025121.5-723433 & sdBN0.2VI:He29\,\dag & 304 & J141420.5-365341 & sdB1I:He12\\
61 & J040304.5-400941 & sdBC1III:He30\,\dag & 305 & J141458.6-461719\,* & sdBC4IV:He40\\
64 & J041319.4-134102 & sdOC7.5VII:He39\,\dag & 313 & J142549.7-043233 & sdO8.5VII:He40\,\dag\\
67 & J042034.8+012040 & sdOC9VII:He39\,\dag & 314 & J142602.3-730649 & sdO4V:He03\\
69 & J042910.7-290248 & sdO8.5VI:He39\,\dag & 336 & J150652.0-464256 & sdO9.5VII:He38\\
73 & J043413.5-423915 & sdO4VI:He02 & 340 & J151417.0-111412 & sdON9.5VI:He34\\
77 & J044208.3-182054 & sdO8VI:He39 & 345 & J152332.3-181726 & sdOC9VI:He39\,\dag\\
79 & J045332.1-370142 & sdB0.5VI:He40\,\dag & 347 & J152607.1-392919 & sdB0.5V:He10\\
81 & J045918.8-535254 & sdB0.5VI:He40 & 348 & J152802.6-333413 & sdB1III:He11\\
85 & J051443.9-084806 & sdB0V:He19 & 354 & J153859.3-483557\,* & sdBC6II:He38\\
90 & J052612.3-285824 & sdO8.5VII:He40\,\dag & 356 & J154033.3-044812 & sdOC2VII:He40\,\ddag\\
92 & J053058.0-684509 & sdO4VI:He03 & 365 & J154816.7-325256 & sdOC9VI:He39\\
95 & J054154.5-305733 & sdO9VII:He14 & 366 & J155036.4-665208 & sdOC2V:He38\\
112 & J061237.4-271255 & sdB0.2VI:He39\,\dag & 373 & J160011.8-643330 & sdB0.5VII:He21\\
114 & J062011.5-583904 & sdO2VI:He06 & 387 & J162835.4-091931\,* & sdBC0II:He40\\
117 & J062709.3-240254 & sdOC5IV:He40 & 393 & J163828.3-245111 & sdO9VI:He27\\
128 & J070738.7-622241 & sdOC7VI:He40\,\dag & 394 & J163947.4-663151 & sdON9.5VI:He36\\
131 & J071436.6-323229 & sdO2V:He09 & 405 & J165955.5-311757 & sdB0.5VII:He06\\
133 & J071656.4-281944 & sdB0VII:He21 & 409 & J170505.9-715608 & sdOC6VII:He40\,\dag\\
140 & J073719.5-251303 & sdO3VI:He03 & 444 & J174151.6-175348\,* & sdOC9.5II:He40\,\ddag\\
141 & J073846.4-193141 & sdO4VI:He03 & 446 & J174234.0-465848\,* & sdBC0.5II:He36\\
143 & J074751.5-253404 & sdB8VII:He03 & 448 & J174407.6-523150 & sdOC2V:He39\\
146 & J075302.4-274958 & sdO8.5VII:He40 & 462 & J180218.6-131029 & sdO5VI:He02\\
148 & J075448.2+014612 & sdB1VII:He19 & 466 & J180655.5+062157\,* & sdBC1V:He39\,\ddag\\
155 & J075807.5-043205 & sdO9.5VII:He32\,\dag & 469 & J180818.5-650853 & sdON9.5VII:He38\\
159 & J080230.8-473606 & sdO4V:He05 & 474 & J181356.8-652747 & sdOC2V:He40\\
162 & J080950.3-364740 & sdO4VIII:He10 & 476 & J181615.3-410325 & sdB3III:He15\\
166 & J082328.9-165019 & sdOC2IV:He38 & 479 & J182314.7-563744\,* & sdBC3II:He40\\
170 & J083523.9-015552 & sdO8VII:He40 & 489 & J183345.9-442230 & sdO4VI:He01\\
172 & J084126.0-350354 & sdO9VI:He11 & 491 & J183349.4+074008 & sdO3VI:He03\\
173 & J084528.8-121409 & sdOC9.5V:He39\,\dag & 495 & J183659.5-462201 & sdO3VII:He02\\
177 & J085158.9-171238 & sdB0.5VI:He40 & 496 & J183716.8-312515 & sdO8.5VI:He39\,\dag\\
178 & J085852.6-131422 & sdB3V:He05 & 500 & J183845.6-540934 & sdBN0VII:He39\,\dag\\
180 & J090505.4+053301 & sdB0VII:He39\,\ddag & 525 & J185522.2-090725 & sdB2.5III:He18\\
181 & J090708.1-030613 & sdO7.5VI:He39\,\dag & 529 & J185804.9-091425 & sdO2VI:He38\\
182 & J090910.0-420830 & sdO5VI:He02 & 530 & J185922.4-425003 & sdB2.5III:He13\\
187 & J092434.9-202936 & sdO9.5VII:He39 & 532 & J190036.8-252426 & sdO7.5VI:He39\\
188 & J094651.4-380321 & sdB7V:He04 & 536 & J190555.5-443840 & sdOC8.5VI:He40\,\dag\\
191 & J095256.8-371941 & sdO8.5VI:He39 & 541 & J191049.5-441713\,* & sdBC0.2VI:He39\,\dag\\
193 & J100042.6-120559 & sdO8VII:He39\,\dag & 542 & J191109.3-140653 & sdOC6.5VI:He40\,\dag\\
198 & J101421.1-565330 & sdO4VIII:He08 & 543 & J191223.6-624632 & sdBC0VII:He27\\
203 & J103254.3-290755 & sdO3VI:He08 & 547 & J191434.9-615408 & sdB2II:He17\\
210 & J105230.8-353913 & sdBN0VI:He28 & 549 & J191504.1-423504 & sdO8VI:He40\,\dag\\
218 & J112610.8-200138 & sdOC2VI:He40\,\dag & 555 & J191849.8-310440 & sdB3VII:He06\\
219 & J113003.7+013737 & sdOC9.5VII:He39\,\ddag & 561 & J192929.4-441245 & sdB5V:He10\\
220 & J113014.6-471054 & sdB3IV:He19 & 562 & J193010.8-373933 & sdO9VII:He14\\
222 & J113910.2-252055 & sdO9VI:He30 & 563 & J193046.1-304959 & sdOC5IV:He38\\
231 & J122227.6-381423 & sdO3V:He12 & 567 & J193323.7-234553\,* & sdBC0.5VI:He38\,\dag\\
239 & J123323.2+062517 & sdON9.5VII:He39\,\ddag & 579 & J195522.1-353437 & sdB3IV:He19\\
240 & J123359.4-674928 & sdO6VII:He36 & 581 & J195630.7-442218\,* & sdB3IV:He35\,\dag\\
241 & J123406.0-344532 & sdOC6VI:He39 & 586 & J200439.2-424627 & sdO3VII:He03\\
242 & J123734.6-284101 & sdO8VII:He40\,\dag & 593 & J201318.9-120118 & sdOC2V:He37\,\dag\\
245 & J124441.6-274858 & sdOC4VII:He40\,\dag & 594 & J201422.9-371542 & sdO9VII:He33\,\dag\\
246 & J124739.2-505317 & sdO5VI:He01 & 598 & J202026.1-190150 & sdO9.5VI:He35\,\dag\\
248 & J124959.9-654427 & sdO5VII:He02 & 600 & J202139.1-342546 & sdO9.5VI:He28\,\dag\\

\hline
\end{tabular}\\
\end{table*}
\addtocounter{table}{-1}
\begin{table*}
\caption[]{ continued.}
    \begin{tabular}{c L{4cm} L{4cm} c L{4cm} l }
\hline
\# & SALT & Drilling class & \# & SALT & Drilling class\\
\hline
610 & J203020.2-595039\,* & sdBC1V:He38\,\dag & 655 & J221628.4-002113 & sdO7.5VI:He39\\
613 & J203540.6-074029 & sdB1.5IV:He14 & 659 & J221901.8-412331 & sdOC9VII:He40\,\dag\\
616 & J203900.5-044240 & sdOC9VII:He20 & 661 & J222122.6+052458 & sdB0.5VII:He20\,\dag\\
622 & J204954.3-693630 & sdO7.5VII:He40\,\dag & 665 & J223522.1-502117 & sdOC6.5VI:He39\\
628 & J205738.9-142543 & sdB1VII:He18\,\ddag & 668 & J223735.8-472240 & sdOC6.5VI:He39\\
630 & J210128.5-545311 & sdO9VI:He37 & 671 & J224830.0-681140 & sdO7.5VI:He39\\
633 & J211111.5-480256 & sdOC7.5VII:He39\,\dag & 672 & J225219.8-631554 & sdO7VI:He39\,\dag\\
644 & J215113.1-210704 & sdO7.5VII:He40\,\dag & 675 & J230525.6-404633 & sdOC9.5VII:He40\\
645 & J215759.7-333506 & sdO7.5VI:He39 & 679 & J230808.7-600954 & sdOC9.5V:He39\\
647 & J220152.2-755204 & sdOC7.5VI:He40 & 684 & J231053.8-630325 & sdO7.5VII:He39\,\dag\\
651 & J220839.9-520323 & sdO9VI:He13 & 690 & J231949.3-503348 & sdO9.5VII:He06\\
652 & J221032.3-454056 & sdB0.2VI:He09 & 692 & J232909.8-100606 & sdO7.5VII:He40\,\dag\\

\hline
\end{tabular}\\
\end{table*}
\begin{figure*}
\centering\includegraphics[width=\textwidth]{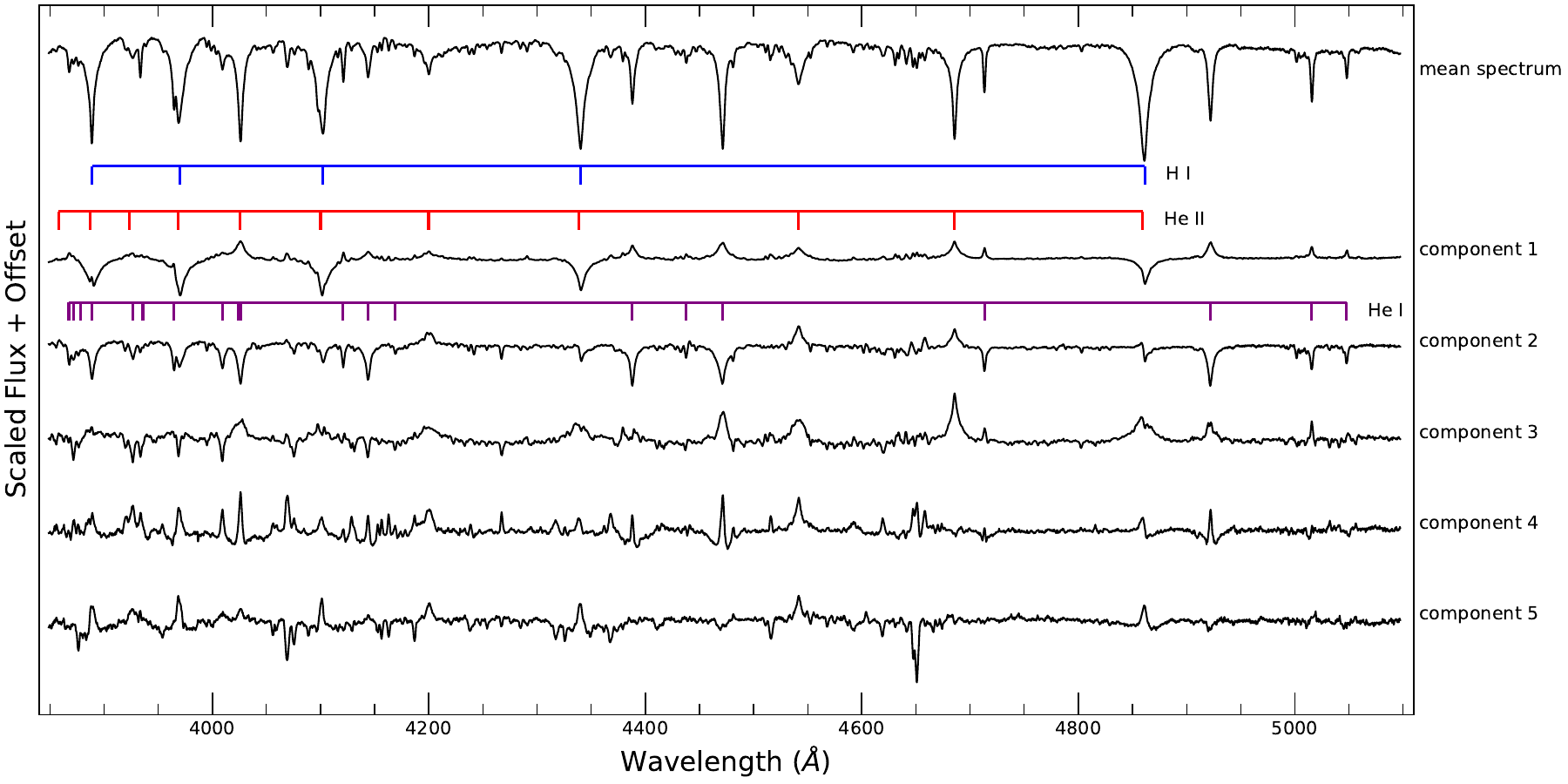}
\caption{ The mean spectrum from 3850 to 5100 \AA\ of the 152 stars used to compute the PC is shown at the top. The first five PC ($u_1,u_2,u_3,u_4,u_5$) spectra are also shown. The PC spectra are arranged in decreasing order of spectral variance. Principal lines are indicated in colour.}
\label{fig:mean_spec}
\end{figure*}

\subsection{Dimensionality reduction -- PCA}
\label{sect:PCA}

PCA is a method of transforming data in a high-dimensional space to a low-dimensional space by representing a set of N-dimensional data using their projections onto a set of optimally defined axes \citep{baron19,winter06.phd}. 
These axes (or principal components - PC) form an orthogonal set of eigenvectors ${\bf E}^{\rm T} = ({\bf u}_1,{\bf u}_2,{\bf u}_3,\ldots $) arranged such that the first PC (${\bf u}_1$) explains the largest variance across the data and each succeeding PC has the highest possible variance.

A standard sample for calculating the PCs was identified using stars with $S/N \geq 100$. This includes 12 Extreme Helium stars (EHes) with $S/N \geq 100$\footnote{The EHe stars are outliers in the SALT sample, having been included because they are extremely H-deficient and may form a continuous sequence with the He-sdB and He-sdO stars.}. The coolest EHes with A-type spectra would complicate the PCs unnecessarily, and have therefore been omitted. The standard sample of 152 stars and their corresponding D13 classifications are provided in Table \ref{t:training}. 
PCs (${\bf u}_i, i=1,152$) were computed from the RSS spectra of these 152 stars using the PCA algorithm from {\sc scikit-learn} \citep{scikit-learn}. 
 
The most significant PCs contain features that are strongly correlated in a majority of the spectra. The first three PCs represent 81\% of the total variance. 
For further analysis, we used the first 14 PCs which  account for 90 per cent of the cumulative variance in the data, providing an optimal choice for rejecting noisy spectra (see also \S \ref{sect:systematic effects}). 

Having computed the PCs for the standard sample ${\bf X}=({\bf x}_1,{\bf x}_2,{\bf x}_3,\ldots )$, where ${\bf x}_i$ is a standard spectrum,  we can compare some unknown spectrum  ${\bf y} = (y_1, y_2, y_3, \ldots )$ to the standard sample by finding its projection ${\bf p}=(p_1,p_2,p_3, \ldots)$ onto the PCs such that  
\begin{equation}
    {\bf p} = ({\bf y} - \overline{\bf x})\cdot\,{\bf E}.
\end{equation}
$\overline{\bf x}$ is the mean spectrum of ${\bf X}$ and, since ${\bf E}$ is orthogonal, ${\bf E } = ({\bf E}^{\rm T})^{\rm T}$ represents the transpose of the first 14 PCs. 
A reduced reconstruction (${\bf y}_r$) of the spectrum can be obtained from, 
\begin{equation}
    {\bf y}_{r} = \overline{\bf x} + {\bf p}\cdot\,{\bf E}^{\rm T}
\end{equation} 
The projection allows for gaps in the sample spectra caused by cosmic ray events or partial observations obtained at only one grating angle. 

This step also acts as a filter for subsequent analysis.
Spectra which cannot be satisfactorily reconstructed are likely to be noisy or otherwise poorly represented by the PCs and should not be used. 
We define the RMS error between an original spectrum and its reduced reconstruction \citep{winter06.phd},
\begin{equation}
    R = \sqrt{\frac{1}{N} \sum_{j=1}^{N} (y_{j} - y_{r,j)^2}}
\label{eq:RMS}
\end{equation}
where $y_j$ is the $j^{th}$ flux bin of ${\bf y}$ and $y_{r,j}$ is the $j^{th}$ flux bin of ${\bf y}_r$.
The mean reconstruction error for the full sample $\bar{R} = 2.54\pm1.69$. 
We identify noisy spectra as those in which $R$ exceeds $2\bar{R}$. 
Here, this accounts for the spectra of 42 stars, 
leaving 587 stars with spectra having  $R \leq 2 \bar{R}$ which are used for further analysis.

The mean spectrum and the first five PCs (${\bf u}_i, i=1,5$) are presented in Fig. \,\ref{fig:mean_spec}. 
It is apparent that the most important spectral features are distributed among multiple PCs. For example, information about the He {\sc ii} lines are spread across at least the first three PCs. 
It is also interesting that each PC contains correlated information about different spectral features. For example, Balmer line features and He {\sc ii} features are seen in the first PC. 
A stellar spectrum is not a linear combination of spectral features and hence a linear decomposition is not be expected to isolate specific features \citep{bailerjones98}. 
However, we can attempt to interpret the features represented by the components.
Component 1 (${\bf u}_1$) is dominated by contributions of H\,{\sc i}, He\,{\sc i} and He\,{\sc ii} lines, with H being anti-correlated with He. 
We will see below that the projections for ${\bf u}_1$ are instrumental in distinguishing the cool H-deficient ($p_1 \ll 0$) from hot H-deficient ($p_1\approx 0$) from H-rich stars ($p_1 \gg 0$). 
Component 2 is again dominated by these major species, with He{\sc ii} being anti-correlated with H and He{\sc i}. This component helps to separate spectra in terms of temperature using the ratios of He{\sc ii} 4686 \AA\ and He{\sc i} 4713 \AA\ line ratios.
The strongest features in component 3 appear to be the dominant He{\sc i} and {He \sc ii} lines, but there are now additional contributions from subordinate lines at all wavelengths. 
Higher order components become increasingly complex.  

\begin{figure*}
\includegraphics[width=0.7\textwidth]{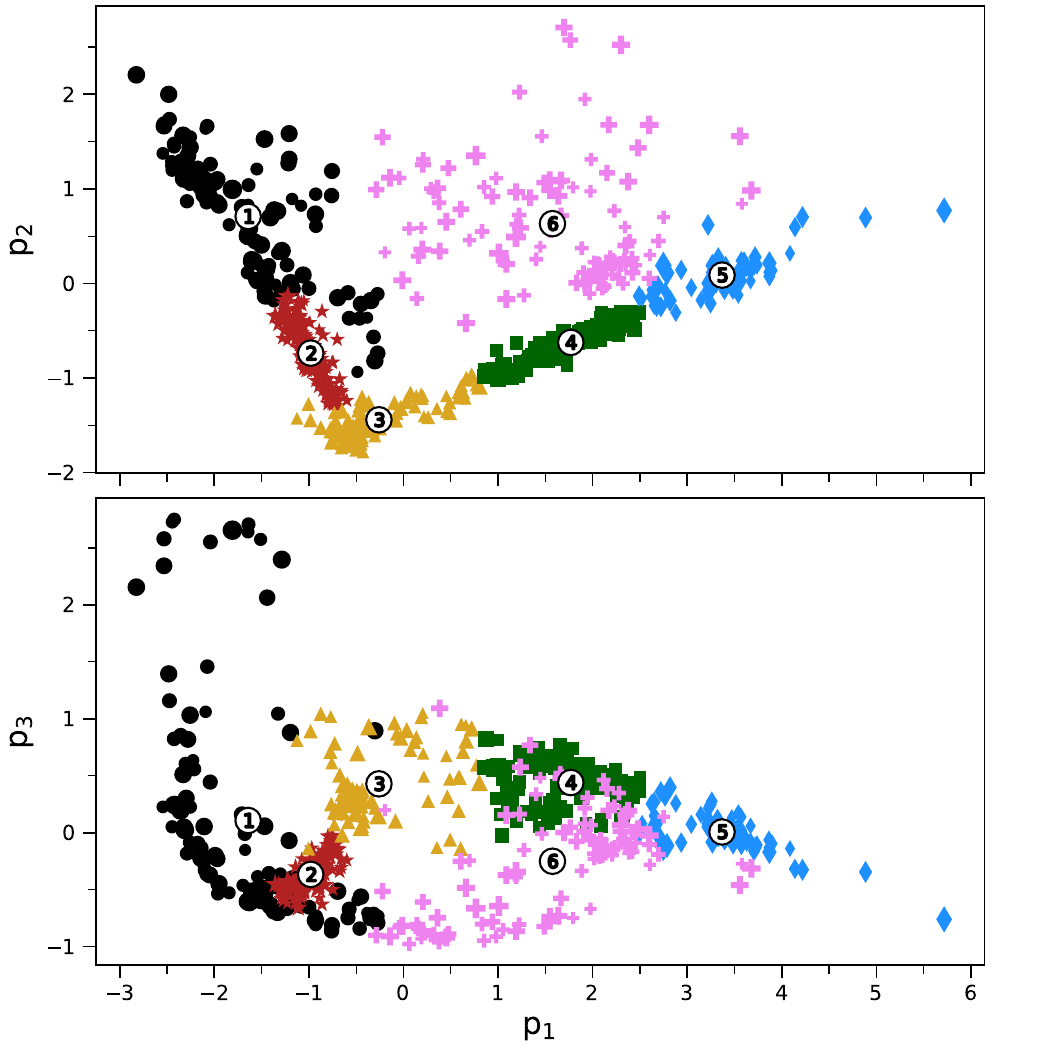}
\caption{Projections $p_1,p_2,p_3$ of the first three PCs ($u_1,u_2,u_3$). 
The clusters obtained from the method described in \S \ref{sect:clustering} are overlaid in different colours with the cluster number indicating the cluster centre. The plot size is proportional to the confidence of the observation obtained as the RMS error described in \S \ref{sect:analysis}. 
}
\label{fig:PC1-3}
\end{figure*}

\subsection{Clustering}
\label{sect:clustering}

Following dimensionality reduction, the dataset can now be processed automatically to identify potential groups. 
Clustering is a family of widely used techniques for exploratory data analysis, of which Spectral Clustering \citep{spielman96} is a good choice for non-uniform data. 
We use the Python code {\sc SpectralClustering} (SC) from {\sc scikit-learn} to apply Spectral Clustering to projections ($p_1,p_2,p_3$) of the first three PCs (Fig. \,\ref{fig:PC1-3}).
Following \citet{luxburg07}, SC constructs a nearest-neighbour similarity graph and computes its Laplacian matrix.
The eigenvectors corresponding to the smallest eigenvalues of the Laplacian matrix are used to reduce complexity of data. 
SC then applies a $k$-means clustering algorithm to identify groups (or clusters). 
Using the Silhouette Coefficient \citep{rousseeuw87}, which computes the mean distances of objects within a cluster and the mean distances of objects to nearest neighbouring cluster, as a measure of the quality of clustering, we identify 6 to be the ideal number of clusters.
At this point every sample member is assigned to a cluster.

Since the clustering algorithm works on the basis of nearest neighbours in 3 dimensional space, some sample members may have significant probabilities of belonging to more than one cluster. 
To identify such edge cases we compute the probability density function of each cluster using a GMM. GMM is a probabilistic model that represents a distribution as a mixture of several gaussians. It implements an expectation-maximisation algorithm which iteratively adjusts the parameters of the gaussians \citep{murphy12}.
Objects within the $20^{\rm th}$ percentile of this distribution are 
marked as uncertain cluster members. 
This exercise left 467 stars assigned to clusters and 120 stars marked as uncertain.
Concurrently, the cluster membership of a star in the uncertain sample can be ascertained as follows. The probability of each uncertain sample object belonging to each cluster is computed from the probability density function. 
Cluster membership is assigned on the basis of highest probability. This was performed on the 120 objects in the uncertain sample; 95 objects were assigned to clusters while 25 showed a similar probability of belonging to 2 or more clusters. These would likely be objects that are true outliers or those that lie on cluster boundaries. 
They were manually assigned to the cluster having the closest sample member already assigned. The PCA space containing the first three components used for classification is shown in Fig. \,\ref{fig:PC1-3} with the clusters overlaid in colour. 

\subsection{Systematic Effects}
\label{sect:systematic effects}

A number of assumption have been used that will contribute to systematic effects. Known limitations include the following.

Errors on radial velocity measurements will affect radial velocity correction leading to small disparities in the position of elemental lines in the spectra of the standard sample.

The standard sample should adequately reflect the overall distribution of stars in any test sample. 
To reduce noise in the PCs that are used for cluster definition, our standard sample consists of stars with the highest S/N. The D13 classifications for the stars were checked to ensure that most of the D13 parameter space is covered. Owing to a combination of low S/N and number fraction, the hydrogen-rich hot sdOs are the only group that are not well represented.

Reduced reconstruction was used to identify spectra sufficiently represented by the PCs, making an optimal sample for further analysis. 
The reconstruction error $R$ and its mean $\bar{R}$ depend on the number of PCs used. 
By including only spectra with $R \leq 2\bar{R}$, the noisiest spectra are excluded from the cluster analysis. 
Increasing the number of PCs for reconstruction results in stars being reconstructed better, ie $R$ decreases, but more quickly than $\bar{R}$. 
In our example, 41 stars would be excluded with 50 PCs, 38 with 14 PCs and 32 with 3 PCs. 
We adopted 14 PCs as an optimal choice for rejecting noisy spectra from the sample to be used for clustering.  

The choice of how many PCs to use for clustering directly affects the number of principal clusters. Using 2 PCs leads to 7 clusters. These are broadly similar to the 6 clusters identified from 3 PCs, except that cluster 4 is divided, and subcluster 1.7 is assigned to cluster 2. 
Using 4 PCs for clustering reduces to 4 clusters after cleaning and would require additional subcluster analysis, as in the case of 3 PCs. In either case, the process appears to be convergent and leads to the identification of a similar number of subclusters with similar boundaries, but different labels. 

The number of clusters is ascertained from an expectation maximisation procedure which computes the Silhouette Coefficient within a range from 2 to 20. The ideal number of clusters would correspond to the peak in the Silhouette Coefficient. 
Here, this is 6.

\section{Analysis}
\label{sect:analysis}
Figure\,\ref{fig:PC1-3} demonstrates that the SALT survey hot subdwarfs are widely distributed across the first three PCs. 

Some clusters are compact ({\it e.g.} cluster 2) and some more dispersed ({\it e.g.} cluster 6). 
The scatter in clusters 1, 3 and 6 shows that these include stars with spectra that are substantially different from the mean spectrum of the cluster. The spectra of stars in the less dispersed clusters 2 and 4 are substantially similar to one another.

\begin{figure*}
    {\includegraphics[width=\textwidth]{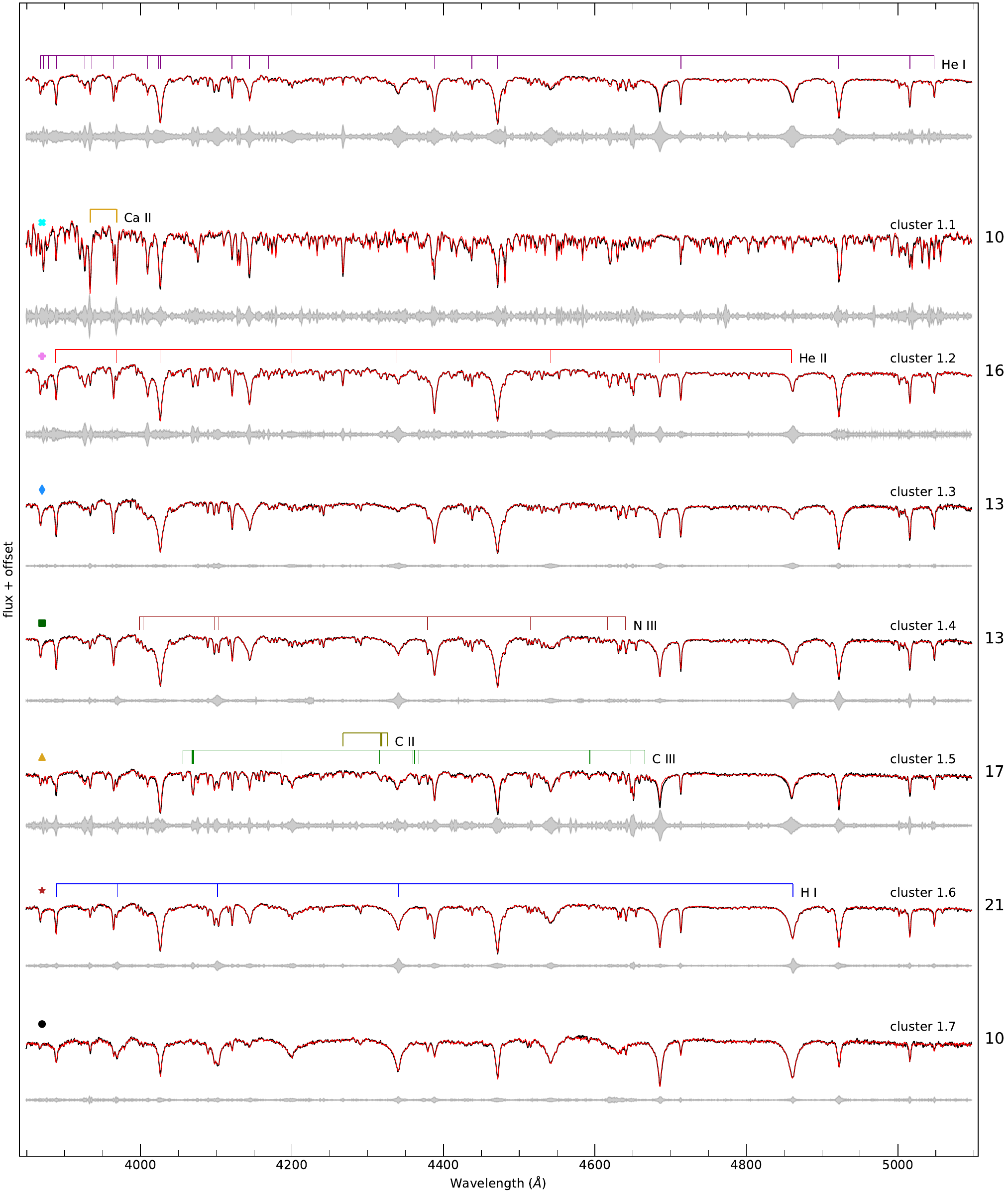}
    \caption{Top: The weighted mean spectrum from 3850 to 5100 \AA\ of cluster 1 - He-sdB/He-sdOB stars.
    Below, in order of decreasing 1st PC: the weighted mean observed spectra (black) and weighted mean reduced reconstruction (red) for each sub-cluster.  
    The sub-cluster variance spectrum (grey) is shown beneath each sub-cluster mean.   
    Each sub-cluster mean spectrum is labelled with the number of members  (right) and the symbol from the corresponding PC diagram (Appendix Fig. \,\ref{fig:cl1PC}). 
    Principal lines are labelled.}
    \label{fig:cl1spec}}
\end{figure*}

\begin{figure*}
    \includegraphics[width=\textwidth]{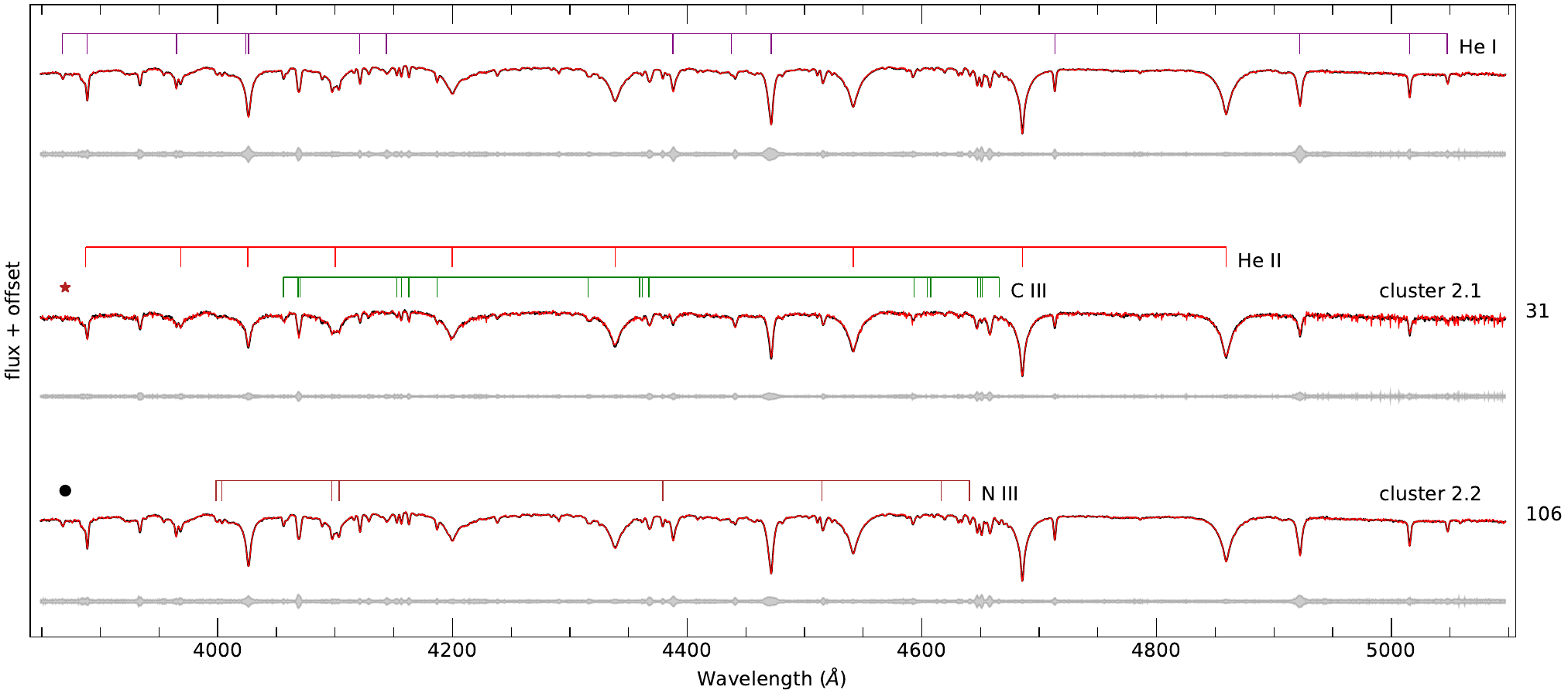}
    \caption{As Fig. \,\ref{fig:cl1spec} for cluster 2 - He-sdO stars. The corresponding PC-diagram is shown in Appendix Fig. \,\ref{fig:cl2PC}.}
    \label{fig:cl2spec}
\end{figure*}

The standard deviation of flux points about the mean spectrum is instrumental in identifying features that vary across the cluster members. These are illustrated in the top panels of Figs.\,\ref{fig:cl1spec} - \ref{fig:cl6spec} which show, for each cluster, the weighted mean observed spectrum, the weighted mean reduced reconstructed spectrum and the standard deviation of the reconstructed spectra about their mean
\footnote{ Weights for each observed spectrum $y_i$ contributing to the mean observed spectrum  $\overline{y_k}$ for each cluster $k$ were taken to be $ w_i = 1 / \sqrt{\sum_{j=1}^{N} (y_{ij} - \overline{y_k})^2 / N}.$
Weights for each reconstructed spectrum $y_{r,i}$ contributing to the mean reconstructed spectrum  $\overline{y_{r,k}}$  were taken to be
the inverse of the reconstruction error given in Eq. \ref{eq:RMS}.}. 
Every mean cluster and sub-cluster spectrum shows Ca{\sc ii} 3933 \AA\ (CaH). With the exception of sub-cluster 1.1, Ca{\sc ii} 3968 \AA\ is always masked by H$\epsilon$ and/or He{\sc ii} 3969 \AA.  
Again with the exception of Cluster 1.1, CaH is of interstellar origin throughout.

Structure within each cluster can be probed by isolating the arbitrary stars in a particular cluster and performing a secondary PCA on the cluster. 
This was necessary because the information contained in the first three PCs of the original sample was insufficient to describe the full dimensionality of the sample.  
We conducted this analysis by creating a cluster sample using all of the spectra identified as belonging to each principal cluster. 
We excluded noisy, incomplete and cosmic ray contaminated spectra, and also all spectra belonging to any other cluster.    
We applied the  methods described already for dimensionality reduction (PCA) and cluster analysis (SC) to each cluster sample. 
Hence we generated a set of secondary PCs for each cluster and identified an optimal number of sub-clusters within each cluster.
The secondary PC and sub-cluster identification diagrams are provided in appendix A.
Here we discuss the clusters and sub-clusters in more detail.



\subsection{Cluster 1 -- He-sdB/He-sdOB}

Cluster 1 is broadly analogous to  the He-sdB and He-sdOB stars. 
We find 7 sub-clusters corresponding to the B-type EHe giants, He-sdBs, He-sdOBs, and a small group of He-sdOs.  
Fig. \,\ref{fig:cl1spec} shows the mean spectrum of the cluster (top) and mean spectra of sub-clusters (subsequently below).
Balmer lines are present in all sub-clusters except 1.1 but H$\beta$ is either weaker than He{\sc i} 4471 (clusters 1.2 -- 1.6) or  weaker than He{\sc ii} 4686 (1.7).
In  clusters 1.2 -- 1.7 the strength of He{\sc ii} 4541 \AA\ relative to H$\beta$+He{\sc ii}4859 \AA\  and H$\gamma$+He{\sc ii}4339 \AA\ indicate the mean spectra to be hydrogen-poor but neither hydrogen-deficient or of intermediate nature (cf. \S\,4.5)
The low-order secondary PCs are strongly sensitive to He{\sc i}, He{\sc ii}, C and N lines.

Sub-cluster 1.1 comprises 10 EHe giant stars. Their mean spectrum is distinguished by the presence of strong stellar calcium H and K lines, indicative of effective temperatures less than 20\,000 K \citep{philipmonai24}.

The next five sub-clusters are arranged in order of increasing He {\sc ii} 4686 \AA\ to He {\sc i} 4713 \AA\ line ratio, which acts as a temperature indicator.

Sub-cluster 1.2, with 16 members, shows a spread in the strength of He {\sc ii} 4686 \AA, C {\sc ii} 4267 \AA\ and C{\sc iii} 4069, 4647 \AA\ lines. 

Sub-clusters 1.3 and 1.4, with 12 and 14 members respectively, are carbon-poor and nitrogen-rich He-sdBs, with the latter showing some contribution from Balmer lines. 
Sub-cluster 1.4 shows He{\sc ii} 4541 \AA\ slightly stronger than 1.3, reflecting the He{\sc ii} 4686/He {\sc i} 4713 ratio and indicative of being slightly hotter. 
With He{\sc ii} 4686/He {\sc i} 4713 $\simge 1$ these clusters would be identified as He-sdOB stars. 
Owing to the strong nitrogen lines, D13 would have identified these as sdBN.    

Sub-clusters 1.5 and 1.6, with 17 and 21 members respectively, contain the carbon-rich and carbon-poor nitrogen-rich He-sdOB stars. He{\sc ii} 4541 \AA\ is more prominent here indicating higher \Teff.
D13 would have identified sub-cluster 1.5 as sdBC and 1.6 as sdBN. 

The PC diagram (Appendix Fig. \ref{fig:cl1PC}) shows two sequences that extend from sub-cluster 1.1 to sub-clusters 1.2 and 1.5. \citet{philipmonai24} shows that EHes can be divided into two luminosity groups. They identify these two groups with  post-merger evolution of double helium white dwarfs (lower luminosity) and post-merger evolution of a carbon-oxygen plus helium white dwarf (higher luminosity). The former contract towards the helium main-sequence to become He-sdOs. The latter evolve through the low gravity hot-sdO regime to eventually become hot white-dwarfs.

Sub-cluster 1.7 contains 10 stars where the ratio of He{\sc ii} 4686 \AA\ to He {\sc i} 4713 \AA\ is similar to the He-sdO stars in cluster 2, but also shows an unidentified broad feature at $\sim 4630$ \AA\ observed in known magnetic hot subdwarfs \citep{pelisoli22,dorsch25}.  It is remarkable that our unsupervised approach is able to identify this distinct group of stars. 

\subsection{Cluster 2 -- He-sdO}

Cluster 2 stars have stronger He {\sc ii} 4686 \AA\ than cluster 1 stars, indicating higher \Teff. 
SC indicates 2 sub-clusters. 
Fig. \,\ref{fig:cl2spec} shows the mean spectrum of the cluster and both sub-clusters.
The variance spectra show He {\sc ii} 4686 \AA\ to be constant but He{\sc i} to vary across the cluster, pointing to a distribution in \Teff.
Inspection of the sub-cluster mean spectra suggests that sub-cluster 2.1 has a higher \Teff, evident from the weaker He{\sc i} lines. Both 2.1 and 2.2 show carbon and nitrogen lines. 
\begin{figure*}
    \includegraphics[width=\textwidth]{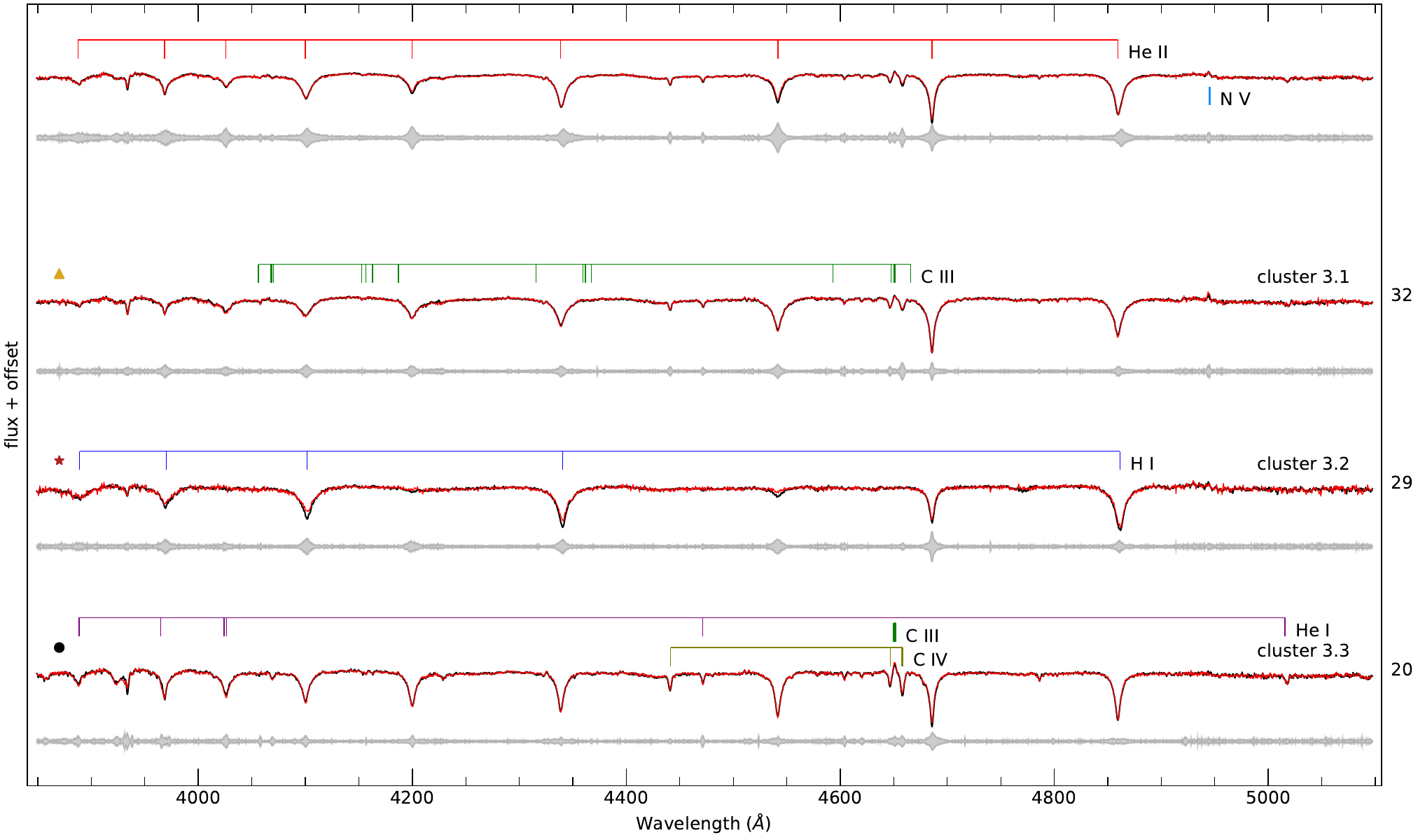}
    \caption{As Fig. \,\ref{fig:cl1spec} for cluster 3 - Hot sdO stars. The corresponding PC-diagram is shown in Appendix Fig. \,\ref{fig:cl3PC}.}
    \label{fig:cl3spec}
\end{figure*}

\subsection{Cluster 3 -- Hot sdO}

Figure\,\ref{fig:cl3spec} shows the mean spectrum of cluster 3. 
The weakness of He {\sc i} 4471 \AA\ indicates very high \Teff, leading us to identify the cluster with hot sdO stars having $\Teff \simge 50$ kK. 
The primary dataset was colour selected and inevitably includes other low luminosity stars including pre-white dwarfs. These stars are not excluded from the sample as they form parts of evolutionary sequences of hot subdwarfs.
The cluster therefore also includes extremely hot DO, O(H) and O(He) stars with $\Teff \simge 100$ kK \citep{jeffery22b}; most of these spectra are noisy and contribute little to the means. 
Secondary analysis identifies 3 sub-clusters including both hydrogen-deficient and hydrogen-rich members (Fig. \,\ref{fig:cl3spec}).  

Sub-cluster 3.1 has 30 members.  
The He{\sc ii} Pickering series is regular and hence H Balmer lines are weak or absent.
The variance spectrum implies a range of hydrogen-helium abundance ratios across the sub-cluster. 
He{\sc i} is weakly visible at 4471 and 5007\AA.
The cluster mean spectrum shows weak C {\sc iii} 4650 \AA\ emission and C {\sc iv} 4441, 4647 and 4658 \AA\ absorption; intra-cluster variation is indicated by the variance spectrum.  
The variance spectrum also indicates that some members show weak N {\sc v} 4945 \AA\ emission. 
This sub-cluster includes 1 O(He) and 3 DO stars.

Sub-cluster 3.2 contains 29 members that show strong Balmer and He{\sc ii} 4686 \AA\  and weak He{\sc ii} 4541 \AA\ absorption. The variance spectrum implies a range of helium/hydrogen abundance ratios across the sub-cluster, but much lower than in cluster 3.1.
This sub-cluster includes four O(H) stars.

Sub-cluster 3.3 has 22 members and the narrowest He {\sc ii} lines indicating the lowest surface gravity within cluster 3. 
The He{\sc ii} Pickering series is regular and hence H Balmer lines are weak or absent. 
The strongest C {\sc iii} 4650 \AA\ emission is seen in this sub-cluster. 
\begin{figure*}
    \includegraphics[width=\textwidth]{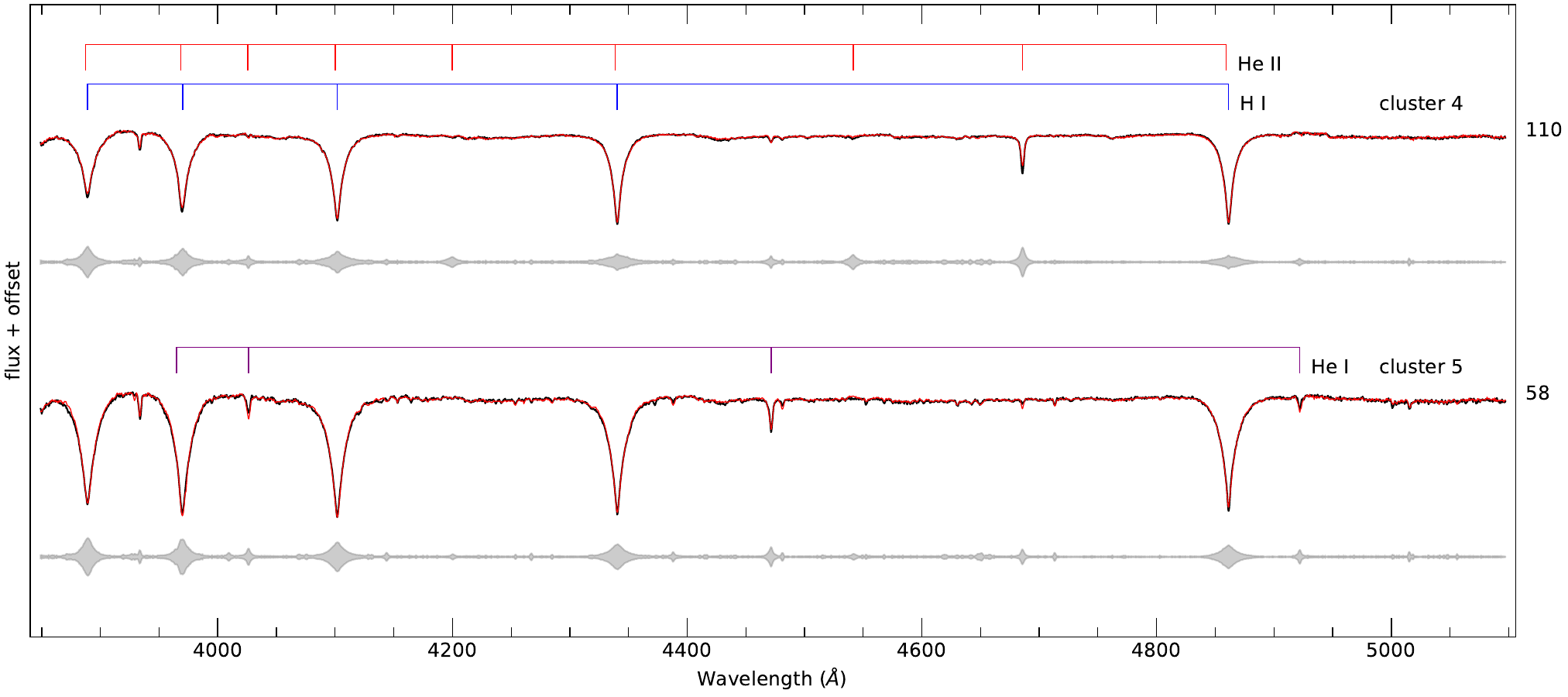}
    \caption{The mean observed and reduced reconstruction spectra for cluster 4 and 5 - classical sdO, sdB stars are shown (top to bottom) as in Fig. \,\ref{fig:cl1spec}.}
    \label{fig:clHspec}
\end{figure*}

\subsection{Clusters 4, 5 -- classical sdO and sdB stars}
Figure\,\ref{fig:clHspec} shows the mean spectrum of clusters 4 and 5. Variance spectra for these two clusters show little dispersion.  
Secondary PCA analysis shows no merit to subdivide into sub-clusters. 
The secondary PC coefficients fall onto straight lines  with no significant dispersion.
In terms of PCs, clusters 4 and 5 form quite distinct groups with very few stars in between. 

Cluster 4 shows Balmer lines and weak He{\sc ii} 4686\AA, corresponding to classical sdO stars. The variance spectrum shows dispersion in the Balmer and He{\sc ii} lines indicating a range in surface gravity and/or temperature and helium abundance.  

Cluster 5 shows Balmer lines with weak He{\sc i} 4026 and 4471 \AA\ lines and corresponds to the classical sdB stars. 
\begin{figure*}
    \includegraphics[width=\textwidth]{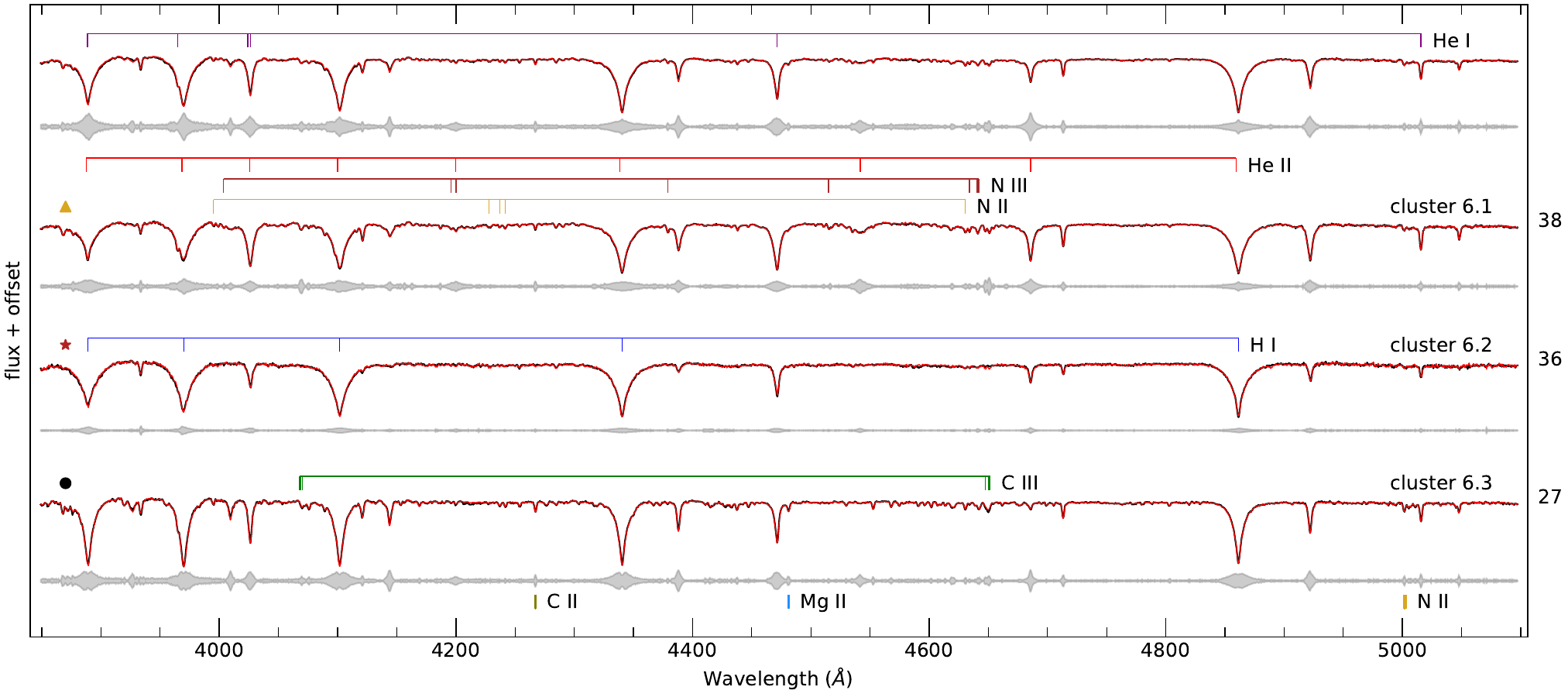}
    \caption{As Fig. \,\ref{fig:cl1spec} for cluster 6 - iHe stars. The corresponding PC-diagram is shown in Appendix Fig. \,\ref{fig:cl6PC}.}
    \label{fig:cl6spec}
\end{figure*}

\subsection{Cluster 6 -- iHe-sdB and iHe-sdOB}

Figure\,\ref{fig:cl6spec} shows the mean spectrum of  cluster 6 and mean spectra of 3 sub-clusters.
All spectra are dominated by Balmer lines, but He{\sc i} lines are much stronger than in sdB stars (cluster 5). 
Consequently these are identified as intermediate helium-rich (iHe) hot subdwarfs.

Sub-cluster 6.1 consists of 37 members with relatively broad Balmer lines.  The ratios of He{\sc ii} 4686 \AA\ and He{\sc i} 4713 \AA\ and the presence of He{\sc ii} 4541 \AA\ indicate this to be the hottest sub-cluster in cluster 6. 
With  He{\sc ii} 4686 \AA\ weaker than He{\sc i} 4471, we apply the label iHe-sdOB.
The spectrum is rich in N {\sc iii} lines including 4196, 4200, 4379, 4515 and 4641 \AA.

Sub-cluster 6.2 consists of 37 members with broad Balmer lines. The He{\sc i} and {\sc ii} lines are weaker than in sub-cluster 6.1, but of roughly the same ratio to one another as in 6.1, so we again apply the label iHe-sdOB.  This sub-cluster is marked by the virtual absence of metal lines.

Sub-cluster 6.3, with 27 members, shows narrower Balmer and He{\sc i} lines than 6.1 and 6.2. He{\sc ii} 4686\AA\ is almost undetectable, making these {\it bona fide} iHe-sdB stars and the coolest sub-cluster in Cluster 6. The mean spectrum shows  C{\sc ii}, C{\sc iii}, N{\sc ii}, N{\sc iii} and Mg{\sc ii} absorption lines. 
\begin{figure}
\includegraphics[width=0.5\textwidth]{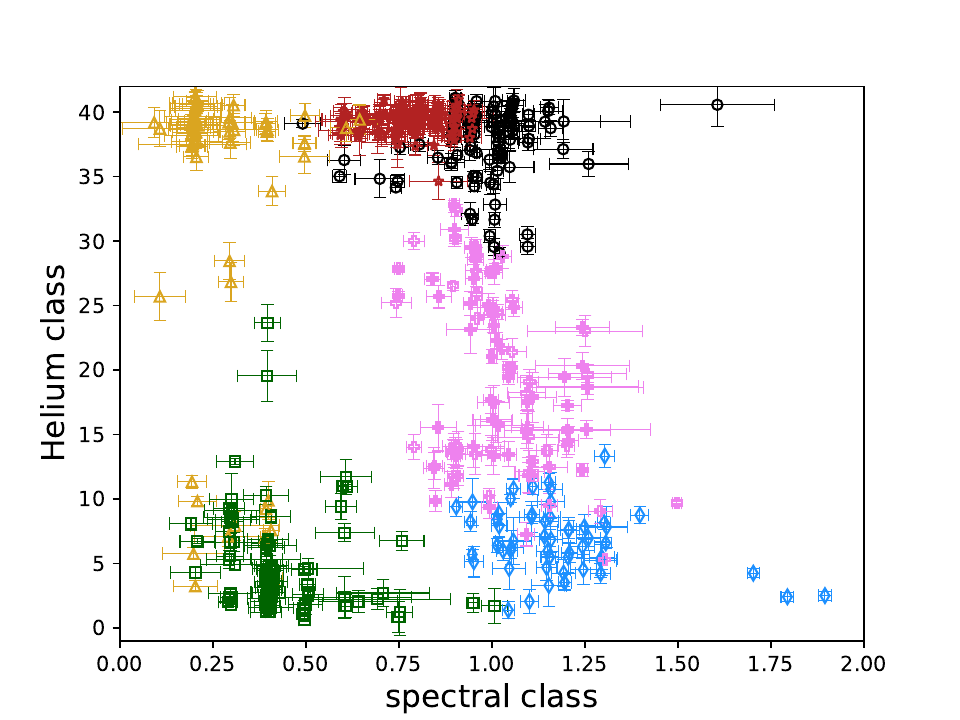}
\caption{The SALT sample plotted using the D13 spectral and helium classes, filtered to exclude stars where the classification error exceeds 0.2 in spectral type or 2 in helium class. Individual stars have had a random dither applied and are coloured according to cluster type 
(\S\ref{sect:clustering}) using the same colours and symbols as Fig. \,\ref{fig:PC1-3}.}
\label{fig:Drilling}
\end{figure}

\subsection{Comparison with D13}

D13 spectral types are available for the full sample using the method described by \citet[][\S 3.1]{jeffery21a} and is available in \citet[][Table A.1]{jeffery25b}.
To compare the clusters derived from PCA with supervised classification, Fig. \,\ref{fig:Drilling} shows the provisional D13 distribution of spectral-type versus helium-class coloured by PCA clusters identified in \S\,\ref{sect:clustering}.  
The clusters are reasonably-well separated in this space, with some overlap at cluster boundaries. 

PCA cluster 1 (He-sdB/He-sdOB/He-sdO) covers an important series of interfaces in D13 classification space. 
At the cool end are EHe stars, likely post-merger products evolving to the helium main-sequence. At the hot end are the cluster 2  He-sdO's, likely helium main-sequence stars. 
Then there are various groups with D13 helium classes 30 - 37 which are more difficult to identify. 
D13 classifications suggest a natural division around helium class 37, but this is not obvious from the PCs,   

One such group includes He-sdO stars with the unidentified 4630 \AA\ feature (sub-cluster 1.7).  
In PC space (Figs.\,\ref{fig:PC1-3}), these form a distinct group ($p_1\approx -0.7$, $p_2\approx -0.8$, $p_3\approx -0.5$) and  show more evidence of hydrogen than the He-sdO stars in Cluster 2.

Our PCA cluster analysis has informed a provisional revision of D13 classification criteria for very hot stars (Jeffery et al. 2025, in prep.).   
D13 defines 9 standards earlier than sdO6.5 of which only 
3 are H-rich, one in each of three luminosity classes.  
Consequently, D13 classifications are ill-defined for these stars.  
Moreover, the D13 line criteria used to define classes break down in this region, since they rely on the depths of He{\sc i} 4471 and He{\sc ii} 4541 which are weak to absent. 

Figure\,\ref{fig:Drilling} shows that cluster 3 stars (hot-sdO) are correctly identified with early spectral types.
The majority are identified with He-rich spectra, but about 1/3 have He-poor spectra (sub-cluster 3.2). 
The latter overlap with cluster 4 stars (sdO), which also have He-poor spectra.   
On average, cluster 3.2 shows slightly stronger He{\sc ii} Pickering lines than cluster 4 (Figs.\,\ref{fig:cl3spec} and \ref{fig:clHspec}).  
For H-rich early sdO stars, the revised classifications rely on the depths of 4 lines (H$\beta$, H$\gamma$, H$\delta$ and He{\sc ii}\,4686: Jeffery et al. 2025, in prep.), while the cluster assignments use PCs defined by the entire spectrum. 
The latter are therefore more sensitive to the Pickering lines, for example, and less sensitive to noise; this difference likely accounts for the overlap on Fig. \,\ref{fig:Drilling}.

The fact that cluster 6 stars (iHe-sdB and iHe-sdOB) are the most spread out in Fig. \,\ref{fig:PC1-3} while the other groups are more tightly bound -- points to the variety of properties within this space and reflects the large spread in helium/hydrogen ratio indicated  by the D13 classifications. 
\begin{table}
    \centering
    \caption{Identification of spectral clusters with spectral classes for hot subdwarfs. Class refers to \citet{moehler90a} with modifications as in \S \ref{sect:intro}. The D13 classes follow from Fig. \,\ref{fig:Drilling}, not accounting for outliers. C and N indicates presence of these lines in an entire sub-cluster. Lines indicated by `+' are found in the majority of cluster members. }
    \begin{tabular}{llll}
    \hline
        \multicolumn{2}{l}{Cluster} & Class &  D13 classes\\
        \hline
         1&  & He-sdB, He-sdOB & sdO9 - sdB3, He 30 - 40\\[1mm]
         &1.1& EHe & \\
         &1.2& He-sdB & sdBC \\
         &1.3& He-sdB & sdBN \\
         &1.4 & He-sdOB & sdO9.5 - sdB0.5 + N lines\\
         &1.5 & He-sdOB & sdO9 - sdB2 + C, N lines\\
         &1.6 & He-sdOB & sdO9 - sdB0 + N lines\\
         &1.7 & He-sdO + 4630\AA & sdO6 - sdO9\\[2mm] 
         2& &He-sdO & sdO6 - sdO9.5, He 34 - 40\\[1mm]
         &2.1& He-sdO & sdO6 - sdO7\\
         &2.2& He-sdO & sdO7 - sdO9.5\\[2mm]
         3& &Hot sdO & sdO2 - sdO6, He 3 - 40\\[1mm]
         &3.1& Hot He-sdO & sdO\\
         &3.2& Hot sdO & sdO, He 3 - 33\\
         &3.3& Hot He-sdO + C{\sc iii} em & sdOe\\[2mm]
         4   & & sdO & sdO1 - sdO9, He 1 - 17\\[2mm]
         5   & & sdB & sdO9 - sdB4, He 2 - 13\\[2mm]
         6   & &iHe-sdB, iHe-sdOB & sdO8 - sdB2, He 6 - 33\\[1mm]
         &6.1 & iHe-sdOB & sdO9.5 + N lines\\
         &6.2 & iHe-sdOB & sdO9.5\\
         &6.3 & iHe-sdB & sdB + C lines\\
         \hline
    \end{tabular}    
    \label{tab:clusters}
\end{table}

\section{Kinematics}
\subsection{Distances}
{\it Gaia} DR3 parallaxes have been used with the zero-point correction described by \citet{lindegren21} to obtain distances. 
Inverting a parallax to obtain distance is only appropriate when errors on measurement are negligible. 
If the fractional parallax error $\sigma/\pi \geq 0.2$, a simple inversion gives an incorrect error estimate. 
This can be mitigated by the use of a properly normalised prior. 
Here an Exponentially Decreasing Space Density Prior is adopted, with one free parameter, the length scale, fixed at 1.35 kpc \citep{luri18}. 
This approach parallels \citet{bailerjones21} in addressing the unreliability of simple parallax inversion for large fractional errors, but differs by adopting a fixed length prior rather than a spatially varying prior.
Of the 587 stars that were identified with clusters in \S\,\ref{sect:clustering}, 57 stars have fractional parallax error $\sigma/\pi \geq 0.2$ of which 20 have $\sigma/\pi \geq 0.3$. Adopting fractional parallax error cut-off $\sigma/\pi = 0.3$, we compute  kinematic parameters for 567 stars. 

RUWE (renormalised unit weight error) is a {\it Gaia} quality control parameter that is related to the chi-square value of the astrometric fit. 
Although a threshold value of 1.4 is recommended \citep{Fabricius21}, \citet{dawson24} show that hot subdwarf stars with companions have RUWE  values up to 5.55 and adopt a more relaxed threshold of 7. Applying this criterion, {\it Gaia} parameters for J074751.5-253405 (RUWE = 20.49) have been discarded.

Additionally, we do not discuss the kinematics for EHes (19 stars) analysed in detail by \citet{philipmonai24}.

172 of the 305 hot subdwarfs identified by \citet{dawson24} from a 500 pc volume-complete sample fall within the sky coverage of SALT. 
Thirteen are helium-rich of which eleven have been observed in the the SALT survey. 
HD 127493 and CD-24 9052 have not been observed with RSS, but have  been discussed elsewhere \citep{kilkenny86,dorsch19}. 
\begin{figure*}
\includegraphics[width=\textwidth]{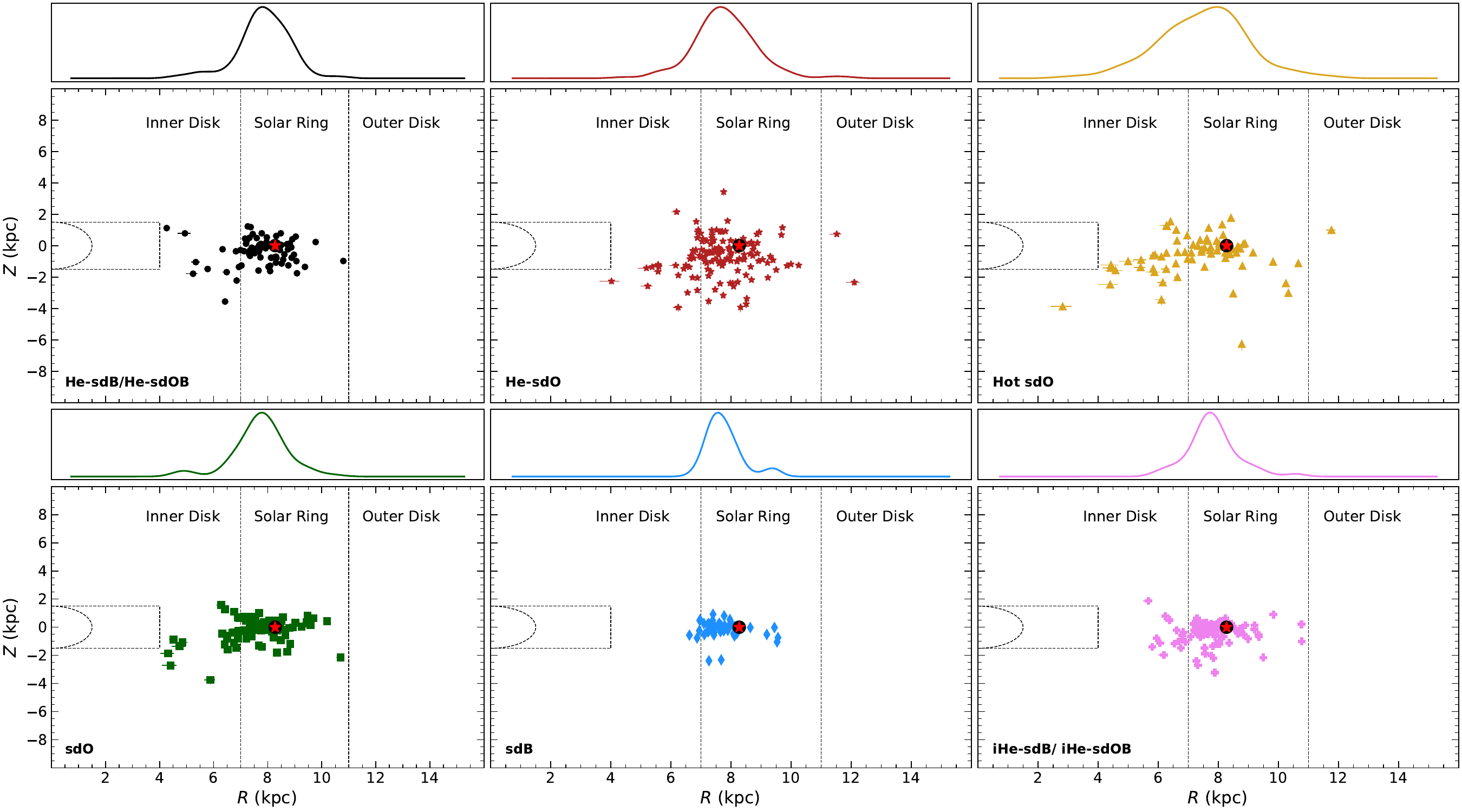}
\caption{Distribution of 547 hot subdwarfs in the $R-Z$ plane for each cluster. Cluster identification is given in the bottom left corner for each panel. The red star denotes the position of the Sun. The dashed ellipse shows the estimates for the inner bulge ($R \sim 1.5$ kpc) and the dashed rectangle shows an estimate of the bar ($X = \pm \,4\, {\rm kpc},\, Y = \pm\,1.5\,{\rm kpc}, Z = \pm\,1.5\,{\rm kpc}$ \citep{galreview16}). The top panel shows the smoothed distribution in $R$. }
\label{fig:RZ}

\includegraphics[width=\textwidth]{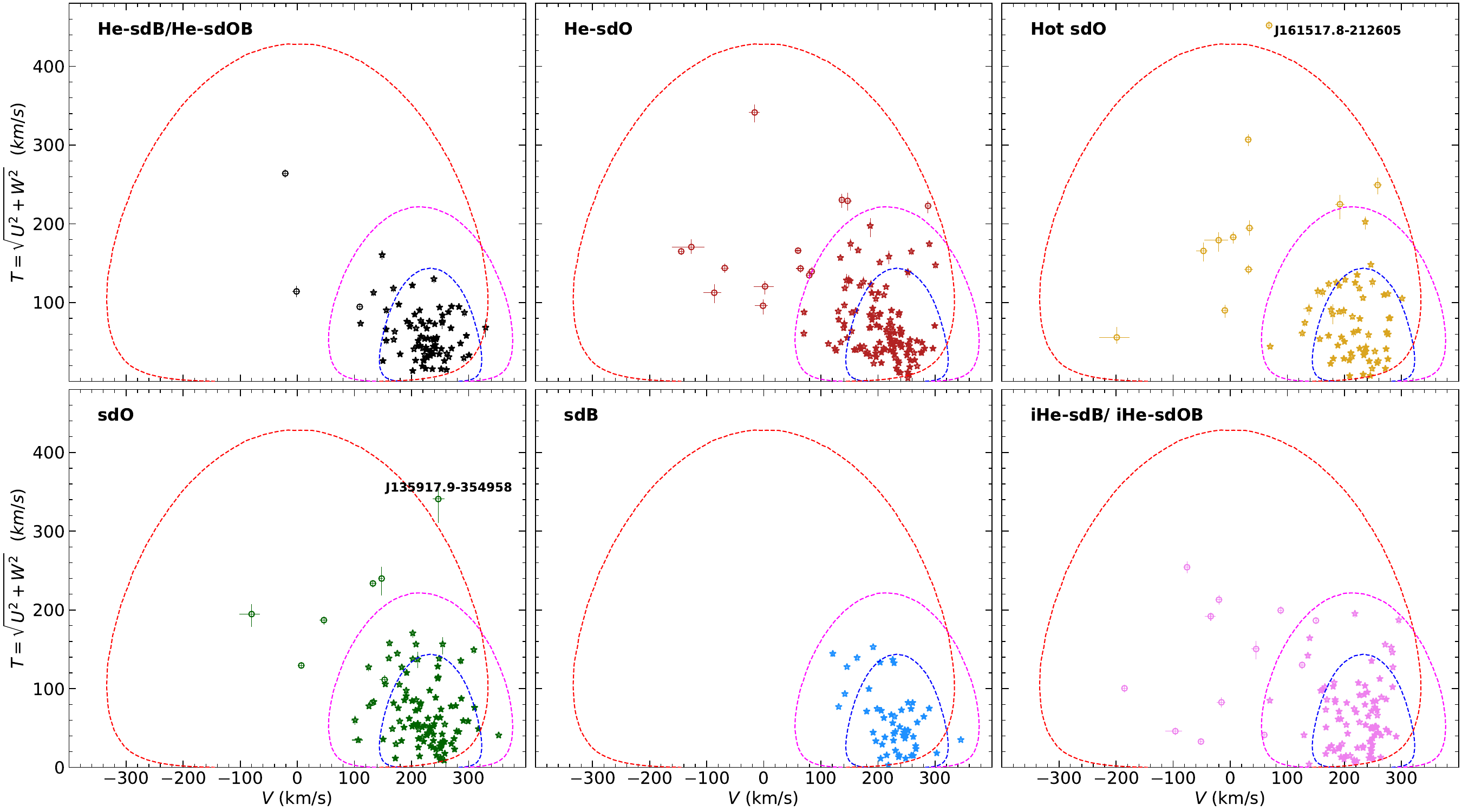}
\caption{The Toomre diagram for the 547 hot subdwarfs: objects are coloured according to cluster type (\S\ref{sect:clustering}) using the same colours as Fig. \ref{fig:PC1-3}. The cluster identification is provided in the top-left corner of each panel. The sample has been classified into disk and spherical components (\S\ref{sect:group_kinematics}). Open circles represent halo objects. Star-shaped symbols represent disk components with filled symbols representing thin disk and open symbols representing thick disk. Dashed lines correspond to $3\,\sigma$ contours from the Besan\c{c}on Galactic models \citep{robin03} for the thin disk (blue), thick disk (magenta) and halo stars (red). J135917.9-354958 and J161517.8-212605 are outliers not falling within the bounds of halo stars.}
\label{fig:Toomre}
\end{figure*}
\begin{figure*}
\includegraphics[width=\textwidth]{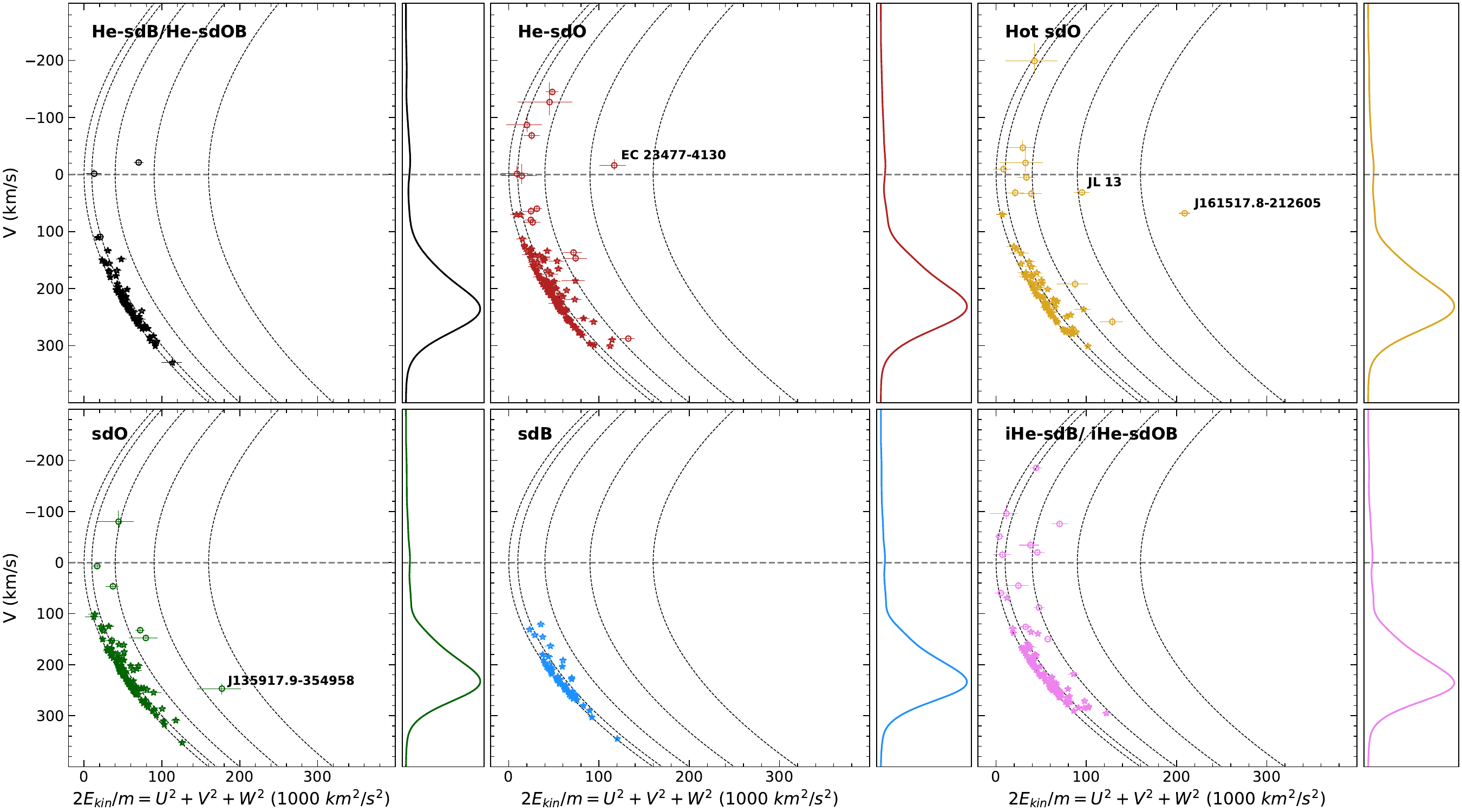}
\caption{Galactic rotational velocity plotted as a function of the total kinetic energy. The colours and symbols have the same meaning as defined in Fig. \ref{fig:Toomre}. The parabolic curves denote lines of equal velocity (($V_{\bot} =(U^{2} + W^{2} )^{1/2}$)) at 0, 100, 200, 300 and 400 \kmsec. The kernel density estimation (right), with binsize 5\kmsec depicts the Galactic rotational velocity distribution for each cluster. The grey dashed line represents $V=0$ with stars above it showing retrograde motion.}
\label{fig:V_Kin}

\includegraphics[width=\textwidth]{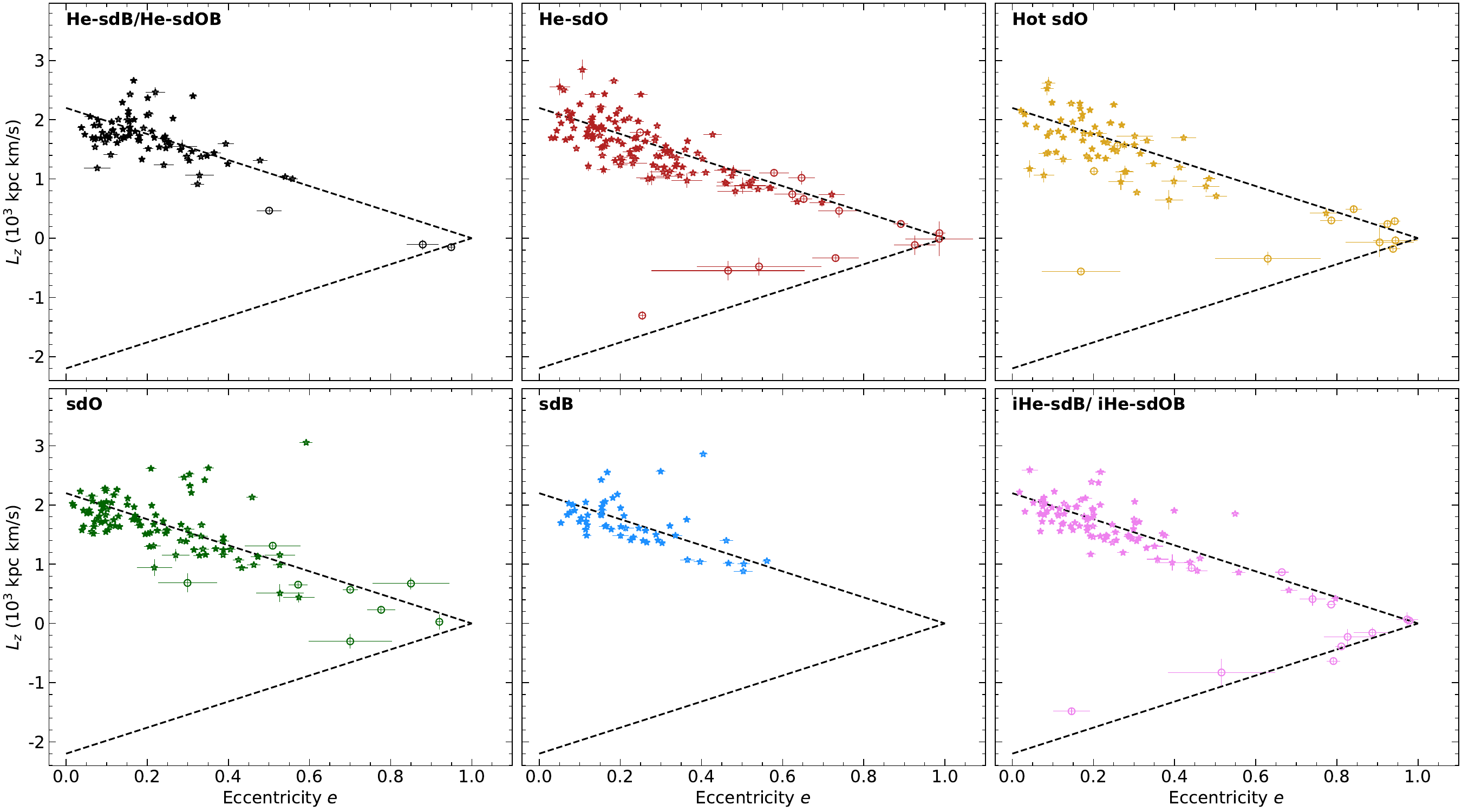}
\caption{The $z$ component of angular momentum $L_z$ plotted as a function of eccentricity $e$. The colours and symbols have the same meaning as defined in Fig. \ref{fig:Toomre}. The dashed lines indicate prograde (positive $L_{\rm z}$) and retrograde (negative $L_{\rm z}$) directions of Galactic rotation.}
\label{fig:ecc_L}
\end{figure*}
\subsection{Galactic and orbital parameters}
Galactic space velocity components were calculated for 547 hot subdwarfs from their radial velocity (\S\ref{sect:SALT}) and the parallax and proper motion from {\it Gaia} DR3 following the method outlined by \citet{randall15}. 
The left-handed system for velocity components is used here, where $U$ is the Galactic radial velocity, positive towards the Galactic Center, V is the Galactic rotational velocity in the direction of the Galactic rotation and W is the component positive towards the North Galactic Pole. 
Other parameters include: distance of Sun to Galactic Centre $8.25 \pm 0.33$ kpc, velocity of the local standard-of-rest relative to the Galactic Center $V_{\rm LSR} = 233.4\pm1.5\kmsec$, and velocity of the Sun relative to the LSR $(v_{\rm x},v_{\rm y},v_{\rm z}) = (9.5,17.3,8.56)\,\kmsec$. 

Figure \ref{fig:RZ} shows the distribution of the stars in the $R-Z$ plane, where $Z$ is the distance from the Galactic plane and $R$ is the Galactocentric distance. 
We define the inner disk $R < 7$ kpc, the Solar ring $7<R<11$ kpc and the outer disk $R>11$ kpc. 
430 of the stars in the sample lie within the solar ring, 114 in the inner disk and 3 in the outer disk.

Members of some clusters (1,2,3, and 4) are observed to much greater distances than, for example, the classical sdBs (cluster 5). 
Since nearly all objects in the sample are brighter than $m_{\rm B}=16$, the former must contain intrinsically brighter stars and hence seen to greater distances.  

The distance distribution of the sample peaks at 1.2 kpc from the Sun and steadily drops; post-asymptotic giant branch stars are observed from $4 - 6$ kpc.

Following \citet{philipmonai24}, we calculate orbits using  \verb|galpy|\footnote{http://github.com/jobovy/galpy} \citep{bovy15}. These are integrated over 5 Gyrs to obtain the Galactic apocentric  and pericentric distances $R_{\rm a}$ and $R_{\rm p}$), the maximum distance from the plane ($Z_{\rm max}$) and the $z$-component of angular momentum ($L_{\rm z}$). The orbital eccentricity is then given by: 
\begin{equation}
    e = \frac{R_{\rm a} - R_{\rm p}}{R_{\rm a} + R_{\rm p}}.
\end{equation}

\subsection{Group kinematics}
\label{sect:group_kinematics}
The definitions of kinematic groups described by \citet{martin17a} and augmented by \citet{philipmonai24} have been used to classify the stars in the sample into those with orbits identified with the Galactic thin and thick disk and halo.
Kinematic population fractions are provided in Table \ref{tab:kin_pop} and are represented graphically in Fig. \ref{fig:Pop_stat}. 

The survey finds relatively few hot subdwarfs in the halo. Since the sample is  limited to $<16^{\rm th}$ magnitude, it detects few hot subdwarfs beyond 3kpc and none beyond 6 kpc; it therefore contains no hot subdwarfs from the Galactic bulge. 
Approximately 2/3 of the He-sdB/He-sdOB, iHe-sdB/iHe-sdOB and sdO stars (clusters 1, 4 and 6) belong to thin disk.  
In contrast, less than half of the He-sdO and hot sdO stars (clusters 2 and 3) belong to the thin disk, with a similar fraction belonging to the thick disk and a significant contribution from the halo (10 -- 14\%).  
86\% of the sdB stars (cluster 5) belong to the thin disk, the remainder are in the thick disk; this group is statistically under-represented since the SALT survey selects {\it against} these stars.   

Figure \ref{fig:Toomre} shows the Toomre diagram for the whole sample. 
The outliers in the Toomre diagram are halo stars with J$135917.9-354958$ having an extended orbit ($R_{\rm a} = 11.69\pm1.37$ ) and J$161517.8-212605$ having a high radial velocity (RV = $272.80\pm9.67$). 
Both have high kinetic energies (Fig. \ref{fig:V_Kin}). 
20 stars exhibit retrograde motion. 
Cluster 5 $-$ sdBs are primarily thin disk stars with a small fraction of thick disk and no halo or bulge stars.

Figure \ref{fig:ecc_L} shows the $z$ component of angular momentum ($L_{\rm z}$) plotted as a function of eccentricity $e$. 
The disk stars are clustered around $e = 0.2\pm0.1$ and $L_{\rm z} = 1700\pm500$, while the halo stars are spread across prograde and retrograde directions of Galactic rotation.

\begin{table}
    \centering
    \caption{ Kinematic population fractions of the clusters. Cluster identification is adopted from \S\ref{sect:analysis}. }
    \resizebox{0.51 \textwidth}{!}{\begin{tabular}{lcccc}
    \hline
        Cluster Identification & N & Thin Disk \% & Thick Disk \% & Halo \% \\
        \hline      
        
        Cluster 1 $-$ He-sdB/He-sdOB &    80 & 66 & 30 & 4\\
        Cluster 2 $-$ He-sdO  &   136 & 44 & 46 & 10\\
        Cluster 3 $-$ Hot sdO &   72  &   46 & 39 & 15\\
        Cluster 4 $-$ sdO &   107 &   66 & 26 & 8\\
        Cluster 5 $-$ sdB &   53  &   83 & 17 & 0\\
        Cluster 6 $-$ iHe-sdB/iHe-sdOB &  99 &   62 & 26 & 12\\ 
        All  & 547 & 59 & 32 & 9 \\ 

        \hline
    \end{tabular}    }
    \label{tab:kin_pop}
\end{table}

\section{Conclusion}
We have developed an unsupervised algorithm for the classification of hot subdwarf spectra obtained during the SALT survey of chemically-peculiar hot subdwarfs. 
The algorithm uses principal components analysis (PCA) for dimensionality reduction and spectral clustering (SC) to define significant classes of spectra. 
We identified 6 major clusters (classes) within the sample; these could be identified with the major classes used to define hot subdwarfs from 10\AA\  resolution spectra.  
A second stage of PCA and SC was used to define new principal components and sub clusters in order to resolve substructure within some of these 6 clusters,
leading to the definition of 17 discrete clusters or sub-clusters. 
The identification of these clusters and sub-clusters with traditional hot subdwarf spectral types is shown in Table\,\ref{tab:clusters}. 

Other results from the unsupervised classification  include the following. \\
1) An ideal training sample would include all spectra available. For the small sample available here it was curated to exclude unrepresentative and noisy data which can seriously disrupt cluster analysis.\\
2) Any representative training set would recover the first three PCs -- which is the basis for our clustering.\\
3) The classification is sensitive to some of the major subtypes identified by \citet{drilling13}, for example, such as the carbon- and nitrogen-rich subtypes. Stars with the unidentified  broad feature at $\sim 4630$ \AA\ are identified and recognised as an independent subgroup.\\
4) PCs can be used to provide relatively clean reconstructions of noisy spectra. In principle  these could provide more precise D13 classifications and help focus follow-up observations for the potentially most interesting stars.  

From an initial sample of 697 stars \citep{jeffery25b}, 587 have been classified autonomously by this method. 
The results are generally in excellent agreement with the distribution of spectral classes identified using a supervised classification scheme based on selected absorption lines \citep{drilling13}. 
We identified a need for better constraints on the classification of sdOs, which are poorly represented by D13 standard stars. 

We also computed the kinematic parameters of 547 hot subdwarfs.
While the full sample shows a majority of thin disk stars (59\%), clusters 2 and 3 (He-sdOs and Hot sdOs), show roughly equal numbers of thin and thick disk stars, as well as a 10\% contribution from the halo. 
Cluster 5 (classical sdBs) has a large fraction (86\%) of thin disk stars and no stars from the bulge or halo.

In the near future, the unsupervised classification algorithm will be projected onto larger datasets. Work has begun on applying the classification to hot subdwarfs in the LAMOST survey \citep{LAMOST12,luo24}. In the future, the project will be extended to include hot subdwarfs from the SDSS and 4MOST surveys \citep{kepler21,4MOST19}.

\section*{Data Availability Statement}
The raw and pipeline reduced SALT observations will be available from the SALT Data Archive (https://ssda.saao.ac.za) after their respective embargo periods. The sky-subtracted, wavelength-calibrated spectra used for this project will be made available from Data Central (https://datacentral.org.au).

\section*{Acknowledgments}
We thank Dr. Matti Dorsch for providing spectral templates for cross-correlation, Dr. Courtney Crawford and the referee for helpful comments.
The Armagh Observatory and Planetarium (AOP) is funded by direct grant from the Northern Ireland Dept for Communities. 
APM acknowledges AOP for a studentship  and the Royal Astronomical Society for a student travel grant to present this work at the 3rd LAMOST-Kepler/TESS workshop.
CSJ acknowledges support from the UK Science and Technology Facilities Council via UKRI Grant No. ST/V000438/1. 

\bibliographystyle{mnras}
\bibliography{ehe} 
\bsp	
\label{lastpage}
\newpage
\appendix
\section{Rejects}
\begin{table*}
    \centering
    \caption{68 stars that were excluded from the \citet{jeffery25b} sample of 697 stars indicated by reason for exclusion as mentioned in \S \ref{sect:SALT}. Stars are indicated by positions (J2000.0) and \# in \citet[][Table A.1]{jeffery25b}.}
    \begin{tabular}{l}
    \hline
    Indication of cool companion from SED\\
        J001006.9-261256 (\# 4), J001853.2-315601 (\# 6), J002334.0+065647 (\# 7), J004905.1-542438 (\# 14), J014844.0-263612 (\# 35), J030434.5-331708 (\# 53), \\
        J031315.1-435459 (\# 55), J031530.1-593404 (\# 56), J042237.3-540849 (\# 68), J043213.8-164508 (\# 72), J043615.2-534334 (\# 74), J051309.7-001220 (\# 83),\\ 
        J055436.0-393722 (\# 99), J062733.6-332222 (\# 118), J064443.6-441402 (\# 121), J065721.8-351137 (\# 126), J071420.9-713815 (\# 129), J075055.6-494310 (\# 144), \\
        J075438.7-280428 (\# 147), J104955.3-271909 (\# 208), J110647.8-572057 (\# 215), J114051.1-225446 (\# 223), J120556.6-110528 (\# 225), J133316.6-074138 (\# 278), \\
        J145858.9-413656 (\# 330), J150206.4-045938 (\# 332), J173035.3-655032 (\# 434), J173827.7-602307 (\# 440), J181508.1+030551 (\# 475), J182259.9-400632 (\# 478), \\
        J183231.8-474437 (\# 487), J184053.9-585548 (\# 504), J191619.4-664738 (\# 551), J191750.2-201409 (\# 554), J201609.3-685333 (\# 595), J202222.2-492940 (\# 601), \\
        J202630.1-624007 (\# 606), J203645.9-251440 (\# 614), J211709.5-700104 (\# 635), J220102.3+083047 (\# 646), J220335.4-733844 (\# 648), J220548.2-351938 (\# 650), \\
        J221656.0-643150 (\# 656), J225635.9-524836 (\# 673), J232417.9-585937 (\# 691) \\[2mm]

    Other classes\\
        J000331.6-164358 (\# 1), J004703.3-115218 (\# 12), J012253.4-752114 (\# 23), J032835.7-205933 (\# 57), J045409.9-592631 (\# 80), J050656.0-251245 (\# 82), \\
        J051949.2-044300 (\# 88), J064133.0+062640 (\# 120), J100133.3-084250 (\# 194), J103958.0-615732 (\# 206), J132534.1-541443 (\# 273), J172411.7-632147 (\# 426), \\ 
        J174009.9-721444 (\# 442), J180707.3-294213 (\# 467),  J181323.6-254640 (\# 472), J182440.9-031959 (\# 481),  J213742.5-382900 (\# 640), J222358.4-251043 (\# 662)\\[2mm]
    
    Incomplete spectra\\        
        J041400.2-315443 (\# 65), J080217.9-004506 (\# 158), J185412.8-542603 (\# 520), J201118.7-315606 (\# 592), J222030.0-020260 (\# 660)\\
    \hline
    \end{tabular}    
    \label{tab:excluded}
\end{table*}

\section{Secondary PC and sub-clusters}
\begin{figure}
    \includegraphics[width=0.51\textwidth]{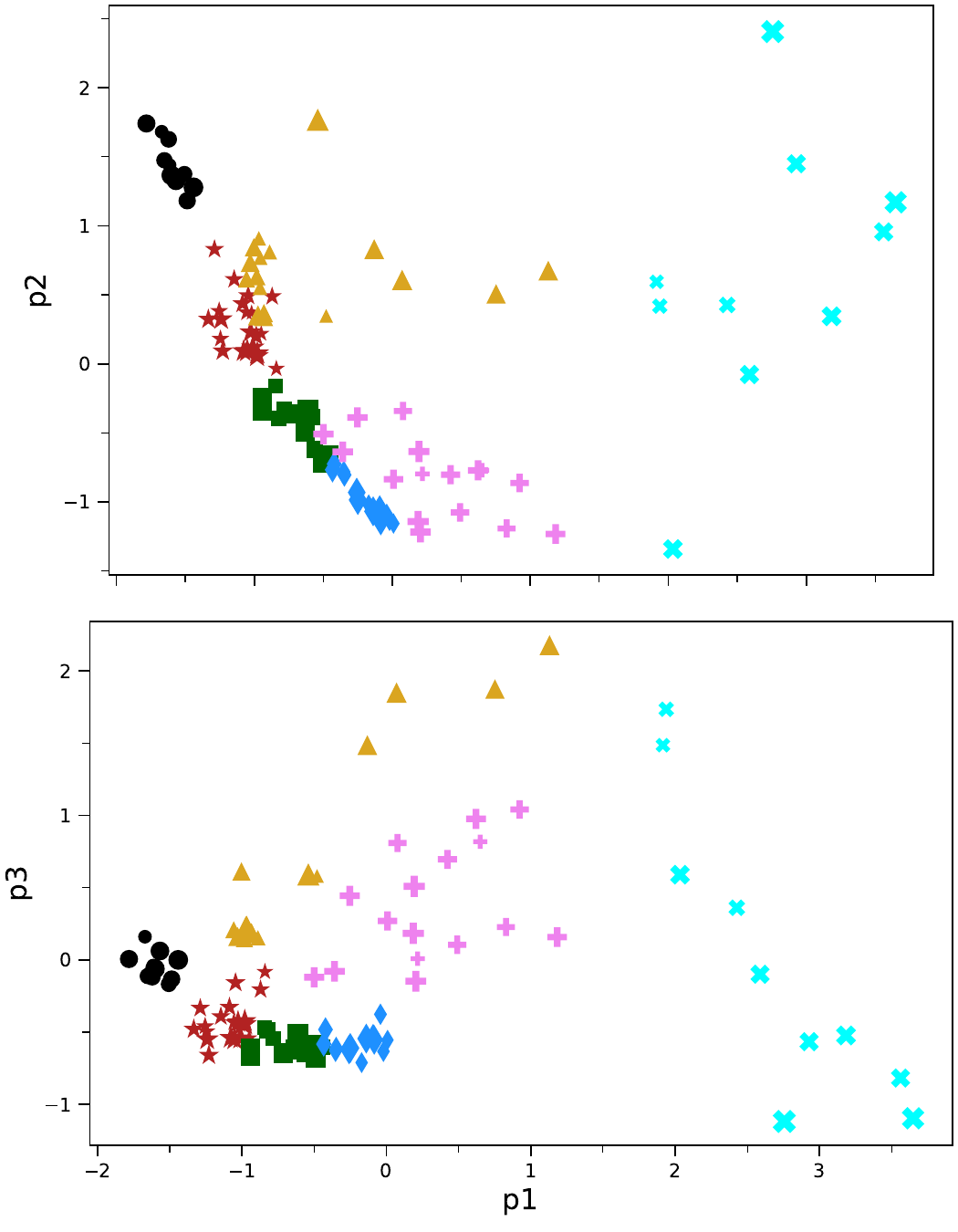}
    \caption{Projections $p_1,p_2,p_3$ of the first three PCs ($u_1,u_2,u_3$) of cluster 1 arranged as in Fig. \ref{fig:PC1-3}}
    \label{fig:cl1PC}
\end{figure}

\begin{figure}
    \includegraphics[width=0.5\textwidth]{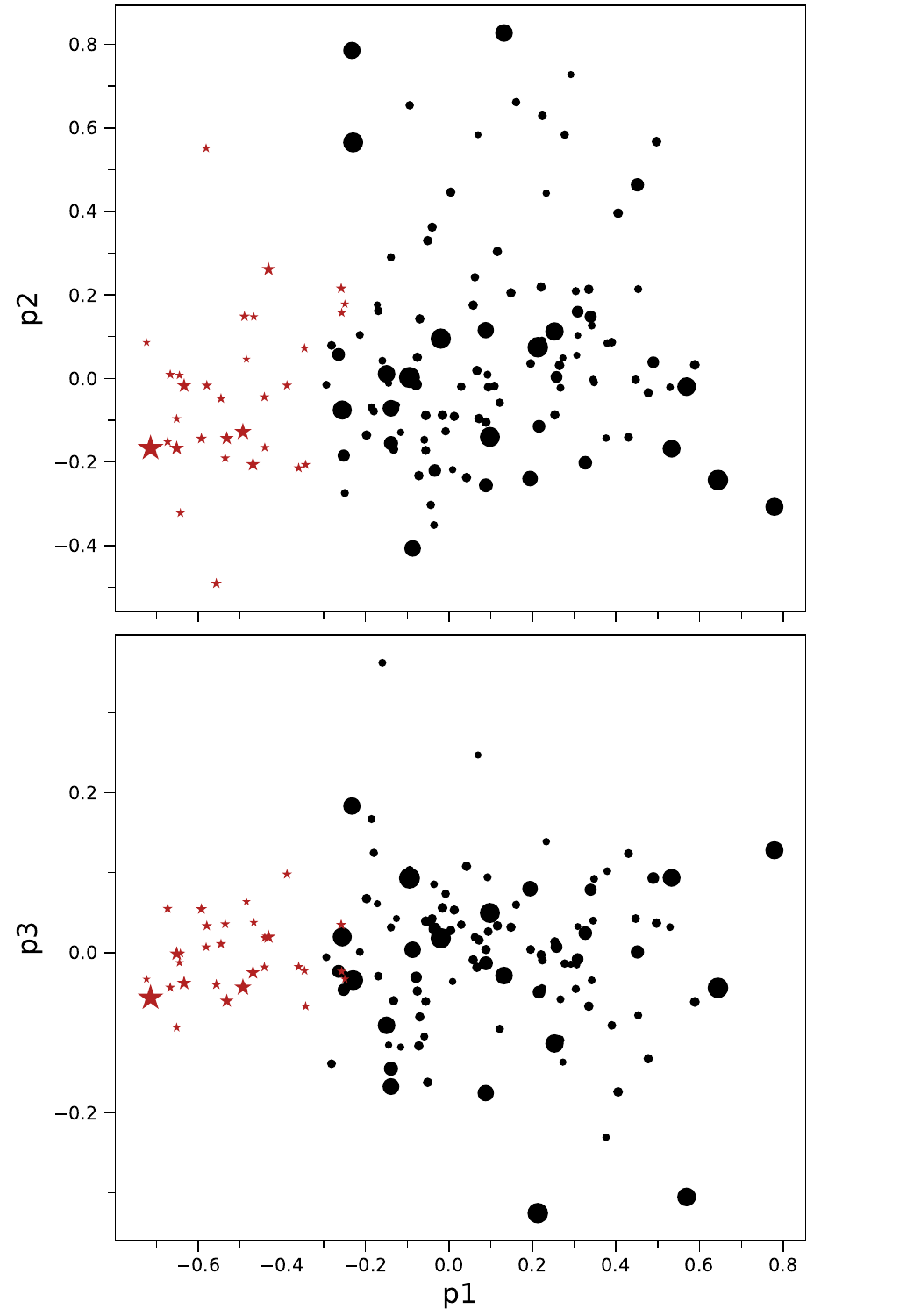}
    \caption{Projections $p_1,p_2,p_3$ of the first three PCs ($u_1,u_2,u_3$) of cluster 2 arranged as in Fig. \ref{fig:PC1-3}}
    \label{fig:cl2PC}
\end{figure}

\begin{figure}
    \includegraphics[width=0.515\textwidth]{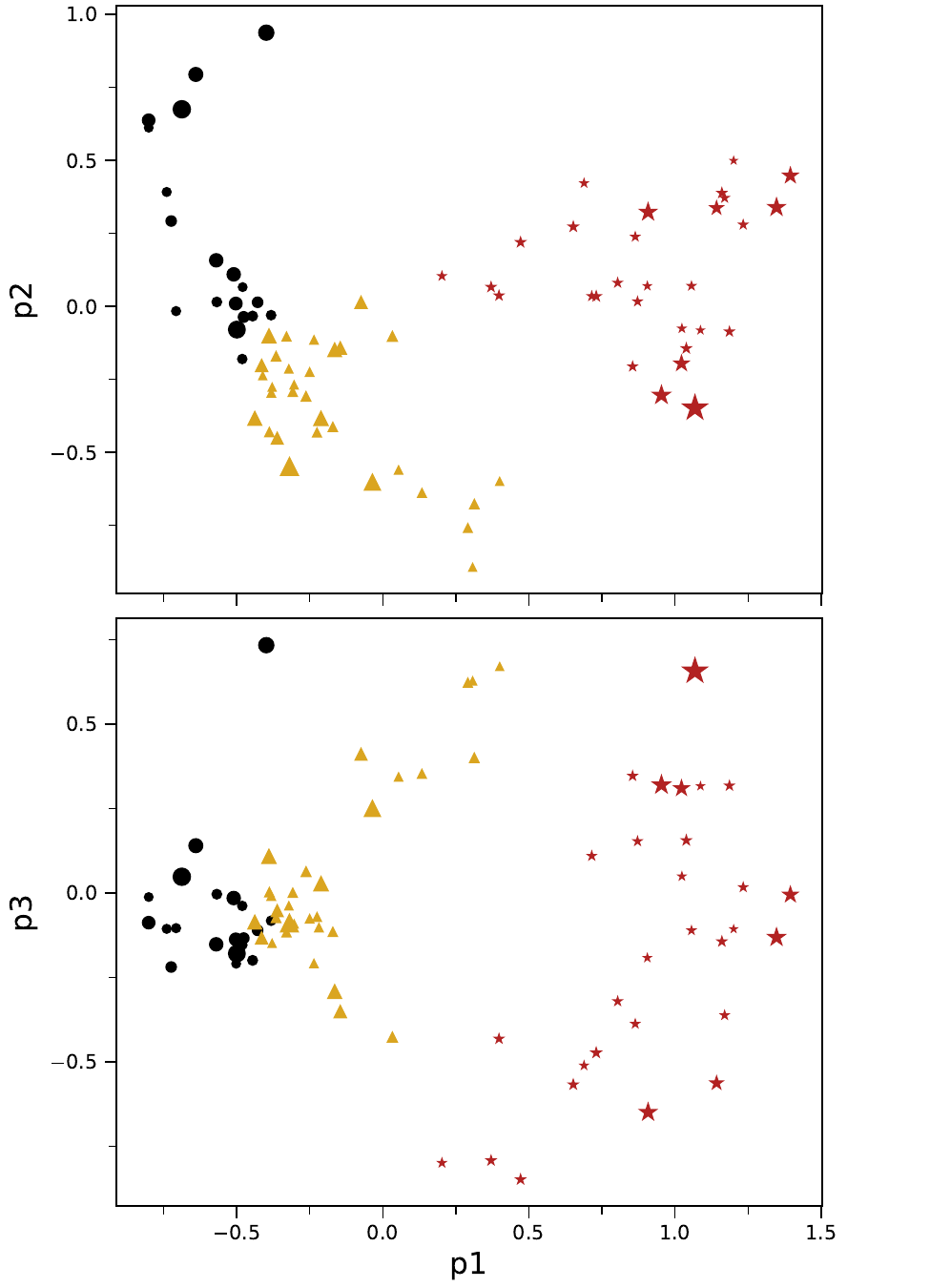}
    \caption{Projections $p_1,p_2,p_3$ of the first three PCs ($u_1,u_2,u_3$) of cluster 3 arranged as in Fig. \ref{fig:PC1-3}}
    \label{fig:cl3PC}
\end{figure}

\begin{figure}
    \includegraphics[width=0.5\textwidth]{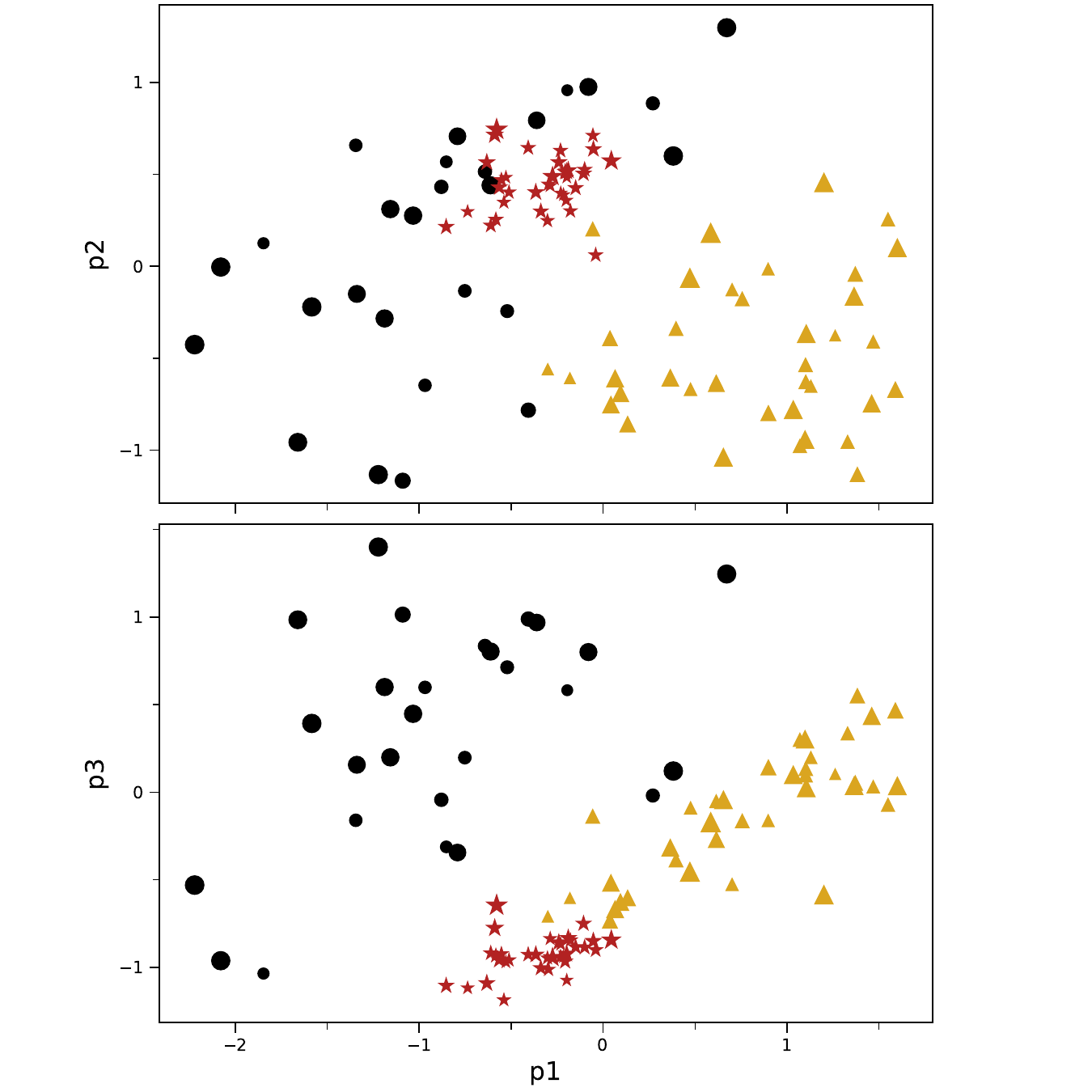}
        \caption{Projections $p_1,p_2,p_3$ of the first three PCs ($u_1,u_2,u_3$) of cluster 6 arranged as in Fig. \ref{fig:PC1-3}}
    \label{fig:cl6PC}
\end{figure}

Figs. \ref{fig:cl1PC}-\ref{fig:cl6PC} show the projections of the first three secondary PCs for clusters 1-3 and 6. The colours and symbols refer to the sub-clusters as indicated along with the weighted mean spectra of sub-clusters in Figs. \ref{fig:cl1spec} - \ref{fig:cl6spec}.


\section{Kinematics}
\begin{figure}
\includegraphics[width=0.5\textwidth]{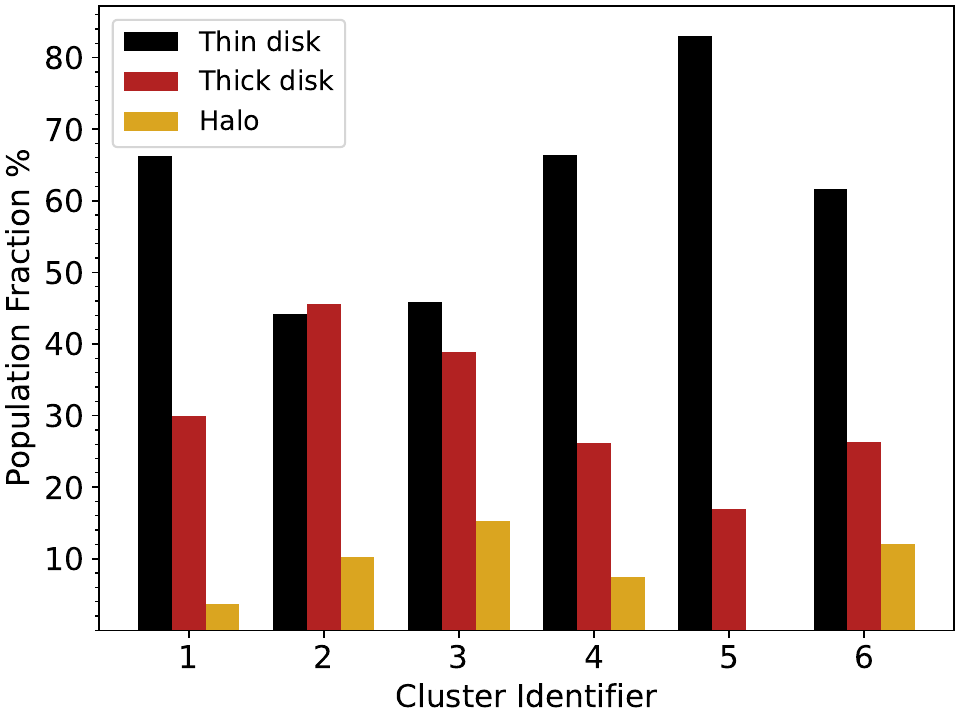}
\caption{Pictorial representation of Table \ref{tab:kin_pop} showing kinematic population fractions for the clusters. Cluster identification is adopted from \S\ref{sect:analysis}. Populations are coloured according to cluster type (\S\ref{sect:clustering}) using the same colours and symbols as Fig. \ref{fig:PC1-3}. Kinematic classification is performed as given in \S\ref{sect:group_kinematics}, dividing the sample into thin disk (TH), thick disk(TK) and Halo (H) populations. Bulge is ignored due to the absence of Bulge population in any of the clusters.}
\label{fig:Pop_stat}
\includegraphics[width=0.5\textwidth]{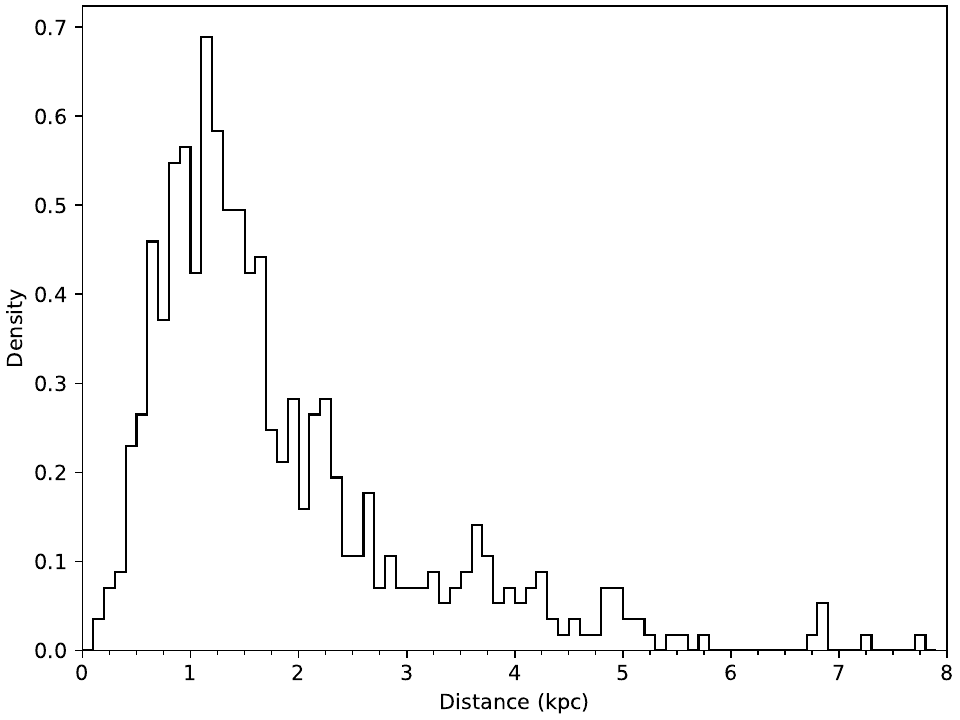}
\caption{ The relative distribution for the 547 hot subdwarfs in Fig.\ref{fig:Pop_stat} and 19 EHes from the SALT sample by distance computed from {\it Gaia} parallaxes. }
\label{fig:Dist_hist}
\includegraphics[width=0.5\textwidth]{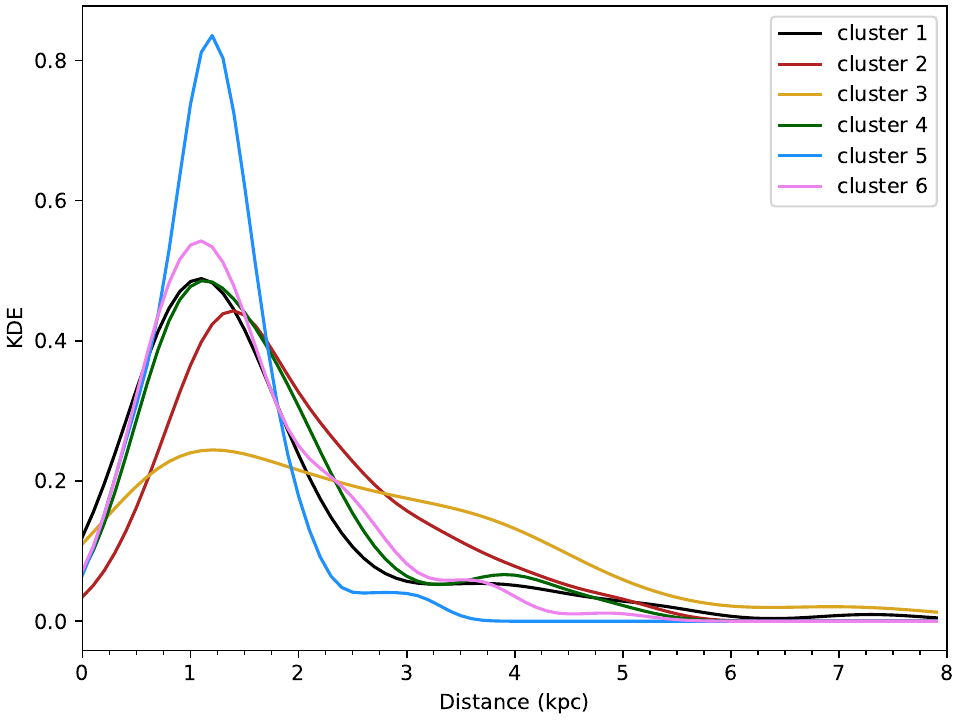}
\caption{Kernel density estimate (KDE) of distance distribution for stars in Fig.\ref{fig:Pop_stat} by PCA cluster assignment.}
\label{fig:Dist_hist_cluster}
\end{figure}

\pagestyle{empty}
\begin{landscape}
\begin{table}
\begin{center}
\caption[]{Computed distance, radial velocities, Galactic velocities and orbital parameters computed for the 547 stars for which kinematic analysis was performed. These parameters have been calculated as mentioned in \S \ref{sect:group_kinematics}. The first column indicates star \# from \citet[][table A.1]{jeffery25b} and second column gives the position (J2000.0). The last column shows the population classification where TH = thin disk, TK = thick disk and H = halo. The stars which exhibit retrograde motion have been marked with $\dagger$.} 
\label{tab:kin_param}
\setlength{\tabcolsep}{4pt}
\tabulinesep=2mm
\begin{tabular}{l L{2.8cm} r@{\,}l r@{$\pm$}l r@{\,}l r@{\,}l r@{\,}l r@{$\pm$}l r@{$\pm$}l r@{$\pm$}l r@{$\pm$}l r@{$\pm$}l c}
\hline
\# &SALT & \multicolumn{2}{c}{Distance}& \multicolumn{2}{c}{$v$} &\multicolumn{2}{c}{$U$} & \multicolumn{2}{c}{$V$} & \multicolumn{2}{c}{$W$} & 
    \multicolumn{2}{c}{$L_z$} & \multicolumn{2}{c}{$e$} & \multicolumn{2}{c}{R$_{a}$} & \multicolumn{2}{c}{R$_p$} & \multicolumn{2}{c}{Z$_{\rm max}$} & pop \\
  & & \multicolumn{2}{c}{kpc} & \multicolumn{2}{c}{\kmsec} & \multicolumn{2}{c}{\kmsec} & \multicolumn{2}{c}{\kmsec} & \multicolumn{2}{c}{\kmsec} & \multicolumn{2}{c}{kpc \kmsec} & \multicolumn{2}{c}{ }   & \multicolumn{2}{c}{kpc} & \multicolumn{2}{c}{kpc}  & \multicolumn{2}{c}{kpc} &  \\
\hline
2 & J000346.7-483937 & $3.83$ & $^{+0.67}_{- 0.99 }$ & $-10.46$ & $6.65$ & $-14.0$ & $^{+1.1}_{- 1.2 }$ & $179.5$ & $^{+4.0}_{- 5.6 }$ & $32.2$ & $^{+3.6}_{- 3.5 }$ & $1267.3$ & $92.6$ & $0.23$ & $0.03$ & $8.06$ & $0.06$ & $5.06$ & $0.34$ & $3.57$ & $0.43$ & TK\\
3 & J000431.0-242621 & $0.69$ & $^{+0.02}_{- 0.02 }$ & $7.87$ & $5.21$ & $-23.6$ & $^{+0.5}_{- 0.5 }$ & $150.0$ & $^{+0.9}_{- 1.2 }$ & $-10.9$ & $^{+2.9}_{- 2.9 }$ & $1254.9$ & $17.1$ & $0.40$ & $0.01$ & $8.39$ & $0.01$ & $3.60$ & $0.07$ & $0.72$ & $0.03$ & TH\\
5 & J001235.6-422011 & $2.28$ & $^{+0.24}_{- 0.30 }$ & $56.25$ & $6.19$ & $-39.7$ & $^{+0.9}_{- 1.0 }$ & $59.7$ & $^{+5.3}_{- 6.5 }$ & $-11.2$ & $^{+3.4}_{- 3.3 }$ & $411.0$ & $109.9$ & $0.74$ & $0.03$ & $8.18$ & $0.02$ & $1.22$ & $0.17$ & $2.96$ & $0.40$ & H\\
8 & J002341.7-532017\,$\dagger$ & $2.23$ & $^{+0.24}_{- 0.30 }$ & $96.17$ & $9.11$ & $-79.2$ & $^{+1.4}_{- 1.5 }$ & $-15.4$ & $^{+7.1}_{- 8.3 }$ & $24.1$ & $^{+6.0}_{- 6.3 }$ & $-227.5$ & $137.3$ & $0.83$ & $0.06$ & $8.16$ & $0.01$ & $0.79$ & $0.29$ & $3.18$ & $0.18$ & H\\
\hline
\end{tabular}
\end{center}
\end{table}
\end{landscape}
\addtocounter{table}{-1}
\begin{landscape}
\begin{table}
\begin{center}
\caption[]{ continued. }
\label{}
\setlength{\tabcolsep}{4pt}
\tabulinesep=1.5mm
\begin{tabular}{l L{2.9cm} r@{\,}l r@{$\pm$}l r@{\,}l r@{\,}l r@{\,}l r@{$\pm$}l r@{$\pm$}l r@{$\pm$}l r@{$\pm$}l r@{$\pm$}l c}
\hline
\# & Star & \multicolumn{2}{c}{Distance}& \multicolumn{2}{c}{$RV$} &\multicolumn{2}{c}{$U$} & \multicolumn{2}{c}{$V$} & \multicolumn{2}{c}{$W$} & \multicolumn{2}{c}{$L_z$} & \multicolumn{2}{c}{$e$} & \multicolumn{2}{c}{R$_{a}$} & \multicolumn{2}{c}{R$_p$} & \multicolumn{2}{c}{Z$_{\rm max}$} & pop \\
& & \multicolumn{2}{c}{kpc} & \multicolumn{2}{c}{\kmsec} & \multicolumn{2}{c}{\kmsec} & \multicolumn{2}{c}{\kmsec} & \multicolumn{2}{c}{\kmsec} & \multicolumn{2}{c}{kpc \kmsec} & \multicolumn{2}{c}{ }   & \multicolumn{2}{c}{kpc} & \multicolumn{2}{c}{kpc}  & \multicolumn{2}{c}{kpc} &  \\
\hline
63 & J041235.2-635719 & $1.65$ & $^{+0.08}_{- 0.09 }$ & $56.45$ & $7.66$ & $21.5$ & $^{+0.6}_{- 0.6 }$ & $172.3$ & $^{+3.2}_{- 3.3 }$ & $11.2$ & $^{+2.9}_{- 2.9 }$ & $1462.5$ & $27.2$ & $0.29$ & $0.01$ & $8.49$ & $0.01$ & $4.68$ & $0.15$ & $1.11$ & $0.04$ & TH\\
64 & J041319.4-134102 & $0.71$ & $^{+0.02}_{- 0.02 }$ & $44.92$ & $5.74$ & $35.4$ & $^{+2.2}_{- 2.3 }$ & $196.1$ & $^{+1.1}_{- 1.2 }$ & $-4.0$ & $^{+2.2}_{- 2.2 }$ & $1840.2$ & $10.3$ & $0.16$ & $0.01$ & $8.98$ & $0.02$ & $6.52$ & $0.08$ & $0.45$ & $0.01$ & TH\\
66 & J041530.2-542158 & $1.89$ & $^{+0.13}_{- 0.15 }$ & $20.75$ & $5.68$ & $-11.4$ & $^{+0.6}_{- 0.6 }$ & $236.4$ & $^{+2.3}_{- 2.3 }$ & $-2.8$ & $^{+2.3}_{- 2.3 }$ & $2080.3$ & $20.4$ & $0.05$ & $0.01$ & $9.35$ & $0.17$ & $8.50$ & $0.04$ & $1.38$ & $0.07$ & TK\\
67 & J042034.8+012040 & $0.41$ & $^{+0.01}_{- 0.01 }$ & $57.86$ & $5.67$ & $71.4$ & $^{+2.9}_{- 2.9 }$ & $197.5$ & $^{+0.7}_{- 0.7 }$ & $1.9$ & $^{+1.9}_{- 1.9 }$ & $1853.1$ & $5.8$ & $0.19$ & $0.01$ & $9.39$ & $0.06$ & $6.34$ & $0.07$ & $0.23$ & $0.01$ & TH\\
\hline
\end{tabular}
\end{center}
\end{table}
\end{landscape}
\addtocounter{table}{-1}
\begin{landscape}
\begin{table}
\begin{center}
\caption[]{ continued. }
\label{}
\setlength{\tabcolsep}{4pt}
\tabulinesep=1.5mm
\begin{tabular}{l L{2.9cm} r@{\,}l r@{$\pm$}l r@{\,}l r@{\,}l r@{\,}l r@{$\pm$}l r@{$\pm$}l r@{$\pm$}l r@{$\pm$}l r@{$\pm$}l c}
\hline
\# & Star & \multicolumn{2}{c}{Distance}& \multicolumn{2}{c}{$RV$} &\multicolumn{2}{c}{$U$} & \multicolumn{2}{c}{$V$} & \multicolumn{2}{c}{$W$} & \multicolumn{2}{c}{$L_z$} & \multicolumn{2}{c}{$e$} & \multicolumn{2}{c}{R$_{a}$} & \multicolumn{2}{c}{R$_p$} & \multicolumn{2}{c}{Z$_{\rm max}$} & pop \\
& & \multicolumn{2}{c}{kpc} & \multicolumn{2}{c}{\kmsec} & \multicolumn{2}{c}{\kmsec} & \multicolumn{2}{c}{\kmsec} & \multicolumn{2}{c}{\kmsec} & \multicolumn{2}{c}{kpc \kmsec} & \multicolumn{2}{c}{ }   & \multicolumn{2}{c}{kpc} & \multicolumn{2}{c}{kpc}  & \multicolumn{2}{c}{kpc} &  \\
\hline
127 & J070322.2-340331 & $0.74$ & $^{+0.02}_{- 0.02 }$ & $74.90$ & $7.88$ & $53.2$ & $^{+2.3}_{- 2.3 }$ & $194.7$ & $^{+4.1}_{- 4.1 }$ & $17.6$ & $^{+1.1}_{- 1.1 }$ & $1767.2$ & $34.9$ & $0.21$ & $0.02$ & $9.16$ & $0.02$ & $5.94$ & $0.22$ & $0.34$ & $0.02$ & TH\\
128 & J070738.7-622241 & $1.44$ & $^{+0.05}_{- 0.05 }$ & $10.87$ & $5.33$ & $37.3$ & $^{+0.5}_{- 0.6 }$ & $241.5$ & $^{+2.6}_{- 2.6 }$ & $22.4$ & $^{+1.1}_{- 1.1 }$ & $2044.8$ & $22.9$ & $0.13$ & $0.01$ & $9.85$ & $0.14$ & $7.56$ & $0.09$ & $0.75$ & $0.03$ & TH\\
130 & J071434.8-224559 & $1.32$ & $^{+0.04}_{- 0.05 }$ & $39.26$ & $5.67$ & $-21.1$ & $^{+1.9}_{- 2.0 }$ & $247.6$ & $^{+2.1}_{- 2.2 }$ & $-1.9$ & $^{+0.3}_{- 0.3 }$ & $2216.9$ & $22.1$ & $0.02$ & $0.01$ & $9.43$ & $0.10$ & $9.10$ & $0.12$ & $0.17$ & $0.01$ & TH\\
131 & J071436.6-323229 & $0.65$ & $^{+0.01}_{- 0.01 }$ & $27.79$ & $5.17$ & $13.0$ & $^{+1.3}_{- 1.3 }$ & $271.7$ & $^{+2.4}_{- 2.4 }$ & $9.7$ & $^{+0.5}_{- 0.5 }$ & $2276.6$ & $23.0$ & $0.17$ & $0.01$ & $11.42$ & $0.19$ & $8.15$ & $0.06$ & $0.11$ & $0.01$ & TH\\
\hline
\end{tabular}
\end{center}
\end{table}
\end{landscape}
\addtocounter{table}{-1}
\begin{landscape}
\begin{table}
\begin{center}
\caption[]{ continued. }
\label{}
\setlength{\tabcolsep}{4pt}
\tabulinesep=1.5mm
\begin{tabular}{l L{2.9cm} r@{\,}l r@{$\pm$}l r@{\,}l r@{\,}l r@{\,}l r@{$\pm$}l r@{$\pm$}l r@{$\pm$}l r@{$\pm$}l r@{$\pm$}l c}
\hline
\# & Star & \multicolumn{2}{c}{Distance}& \multicolumn{2}{c}{$RV$} &\multicolumn{2}{c}{$U$} & \multicolumn{2}{c}{$V$} & \multicolumn{2}{c}{$W$} & \multicolumn{2}{c}{$L_z$} & \multicolumn{2}{c}{$e$} & \multicolumn{2}{c}{R$_{a}$} & \multicolumn{2}{c}{R$_p$} & \multicolumn{2}{c}{Z$_{\rm max}$} & pop \\
& & \multicolumn{2}{c}{kpc} & \multicolumn{2}{c}{\kmsec} & \multicolumn{2}{c}{\kmsec} & \multicolumn{2}{c}{\kmsec} & \multicolumn{2}{c}{\kmsec} & \multicolumn{2}{c}{kpc \kmsec} & \multicolumn{2}{c}{ }   & \multicolumn{2}{c}{kpc} & \multicolumn{2}{c}{kpc}  & \multicolumn{2}{c}{kpc} &  \\
\hline
183 & J090943.5-715348 & $1.06$ & $^{+0.03}_{- 0.03 }$ & $0.81$ & $5.08$ & $-28.8$ & $^{+0.6}_{- 0.6 }$ & $253.5$ & $^{+3.0}_{- 3.0 }$ & $7.9$ & $^{+0.9}_{- 0.9 }$ & $1988.4$ & $22.7$ & $0.02$ & $0.01$ & $8.38$ & $0.15$ & $8.06$ & $0.07$ & $0.28$ & $0.01$ & TH\\
184 & J091715.6-035356 & $2.36$ & $^{+0.25}_{- 0.32 }$ & $57.81$ & $6.34$ & $-161.3$ & $^{+5.3}_{- 7.4 }$ & $152.0$ & $^{+3.1}_{- 3.6 }$ & $-67.1$ & $^{+3.5}_{- 4.8 }$ & $1056.4$ & $64.5$ & $0.48$ & $0.02$ & $9.96$ & $0.15$ & $3.55$ & $0.12$ & $4.78$ & $0.75$ & TK\\
185 & J091856.0-570425 & $0.67$ & $^{+0.01}_{- 0.01 }$ & $27.83$ & $5.39$ & $-46.4$ & $^{+0.3}_{- 0.3 }$ & $224.6$ & $^{+3.1}_{- 3.1 }$ & $-9.1$ & $^{+0.3}_{- 0.3 }$ & $1894.5$ & $25.8$ & $0.15$ & $0.01$ & $9.15$ & $0.07$ & $6.73$ & $0.14$ & $0.17$ & $0.01$ & TH\\
186 & J092103.4-580849 & $1.09$ & $^{+0.02}_{- 0.02 }$ & $29.44$ & $5.39$ & $-8.7$ & $^{+0.1}_{- 0.1 }$ & $227.1$ & $^{+3.2}_{- 3.2 }$ & $39.9$ & $^{+0.4}_{- 0.4 }$ & $1811.7$ & $25.8$ & $0.13$ & $0.01$ & $8.56$ & $0.03$ & $6.60$ & $0.17$ & $0.50$ & $0.01$ & TH\\
\hline
\end{tabular}
\end{center}
\end{table}
\end{landscape}
\addtocounter{table}{-1}
\begin{landscape}
\begin{table}
\begin{center}
\caption[]{ continued. }
\label{}
\setlength{\tabcolsep}{4pt}
\tabulinesep=1.5mm
\begin{tabular}{l L{2.9cm} r@{\,}l r@{$\pm$}l r@{\,}l r@{\,}l r@{\,}l r@{$\pm$}l r@{$\pm$}l r@{$\pm$}l r@{$\pm$}l r@{$\pm$}l c}
\hline
\# & Star & \multicolumn{2}{c}{Distance}& \multicolumn{2}{c}{$RV$} &\multicolumn{2}{c}{$U$} & \multicolumn{2}{c}{$V$} & \multicolumn{2}{c}{$W$} & \multicolumn{2}{c}{$L_z$} & \multicolumn{2}{c}{$e$} & \multicolumn{2}{c}{R$_{a}$} & \multicolumn{2}{c}{R$_p$} & \multicolumn{2}{c}{Z$_{\rm max}$} & pop \\
& & \multicolumn{2}{c}{kpc} & \multicolumn{2}{c}{\kmsec} & \multicolumn{2}{c}{\kmsec} & \multicolumn{2}{c}{\kmsec} & \multicolumn{2}{c}{\kmsec} & \multicolumn{2}{c}{kpc \kmsec} & \multicolumn{2}{c}{ }   & \multicolumn{2}{c}{kpc} & \multicolumn{2}{c}{kpc}  & \multicolumn{2}{c}{kpc} &  \\
\hline
236 & J122550.4-354733 & $2.31$ & $^{+0.25}_{- 0.32 }$ & $50.54$ & $5.25$ & $-165.2$ & $^{+4.3}_{- 5.3 }$ & $202.1$ & $^{+3.0}_{- 2.9 }$ & $42.3$ & $^{+1.4}_{- 1.4 }$ & $1255.5$ & $32.6$ & $0.34$ & $0.02$ & $7.78$ & $0.01$ & $3.86$ & $0.14$ & $1.07$ & $0.07$ & TK\\
237 & J122701.9-343648 & $1.21$ & $^{+0.07}_{- 0.07 }$ & $-23.33$ & $5.29$ & $-27.2$ & $^{+1.0}_{- 1.1 }$ & $266.0$ & $^{+2.4}_{- 2.4 }$ & $-3.1$ & $^{+1.4}_{- 1.4 }$ & $2179.8$ & $19.5$ & $0.19$ & $0.01$ & $11.33$ & $0.20$ & $7.66$ & $0.04$ & $0.79$ & $0.03$ & TH\\
238 & J123052.0-604915 & $1.69$ & $^{+0.09}_{- 0.10 }$ & $15.10$ & $5.31$ & $-75.4$ & $^{+1.4}_{- 1.3 }$ & $231.2$ & $^{+2.8}_{- 2.7 }$ & $-10.2$ & $^{+0.3}_{- 0.4 }$ & $1730.8$ & $24.5$ & $0.16$ & $0.01$ & $8.34$ & $0.03$ & $6.09$ & $0.15$ & $0.25$ & $0.01$ & TH\\
239 & J123323.2+062517 & $0.69$ & $^{+0.02}_{- 0.02 }$ & $-16.31$ & $5.38$ & $-54.0$ & $^{+0.7}_{- 0.6 }$ & $296.1$ & $^{+1.2}_{- 1.3 }$ & $21.0$ & $^{+3.2}_{- 3.1 }$ & $2090.1$ & $8.6$ & $0.15$ & $0.01$ & $10.33$ & $0.08$ & $7.59$ & $0.03$ & $0.81$ & $0.02$ & TH\\
\hline
\end{tabular}
\end{center}
\end{table}
\end{landscape}
\addtocounter{table}{-1}
\begin{landscape}
\begin{table}
\begin{center}
\caption[]{ continued. }
\label{}
\setlength{\tabcolsep}{4pt}
\tabulinesep=1.5mm
\begin{tabular}{l L{2.9cm} r@{\,}l r@{$\pm$}l r@{\,}l r@{\,}l r@{\,}l r@{$\pm$}l r@{$\pm$}l r@{$\pm$}l r@{$\pm$}l r@{$\pm$}l c}
\hline
\# & Star & \multicolumn{2}{c}{Distance}& \multicolumn{2}{c}{$RV$} &\multicolumn{2}{c}{$U$} & \multicolumn{2}{c}{$V$} & \multicolumn{2}{c}{$W$} & \multicolumn{2}{c}{$L_z$} & \multicolumn{2}{c}{$e$} & \multicolumn{2}{c}{R$_{a}$} & \multicolumn{2}{c}{R$_p$} & \multicolumn{2}{c}{Z$_{\rm max}$} & pop \\
& & \multicolumn{2}{c}{kpc} & \multicolumn{2}{c}{\kmsec} & \multicolumn{2}{c}{\kmsec} & \multicolumn{2}{c}{\kmsec} & \multicolumn{2}{c}{\kmsec} & \multicolumn{2}{c}{kpc \kmsec} & \multicolumn{2}{c}{ }   & \multicolumn{2}{c}{kpc} & \multicolumn{2}{c}{kpc}  & \multicolumn{2}{c}{kpc} &  \\
\hline
286 & J135131.2-415047 & $2.40$ & $^{+0.28}_{- 0.36 }$ & $-37.62$ & $6.55$ & $-133.1$ & $^{+5.4}_{- 6.9 }$ & $252.5$ & $^{+3.6}_{- 3.5 }$ & $38.3$ & $^{+1.9}_{- 2.1 }$ & $1459.4$ & $50.7$ & $0.22$ & $0.02$ & $7.74$ & $0.10$ & $4.91$ & $0.24$ & $0.90$ & $0.07$ & TH\\
287 & J135237.7-463925 & $1.84$ & $^{+0.13}_{- 0.15 }$ & $-54.73$ & $6.15$ & $-85.1$ & $^{+3.0}_{- 2.9 }$ & $283.1$ & $^{+2.9}_{- 2.8 }$ & $41.4$ & $^{+1.3}_{- 1.2 }$ & $1804.9$ & $32.0$ & $0.16$ & $0.01$ & $8.49$ & $0.18$ & $6.18$ & $0.11$ & $0.57$ & $0.02$ & TH\\
288 & J135659.9-751827 & $1.28$ & $^{+0.04}_{- 0.05 }$ & $-0.60$ & $5.21$ & $-45.6$ & $^{+1.4}_{- 1.5 }$ & $239.3$ & $^{+2.5}_{- 2.5 }$ & $6.8$ & $^{+0.7}_{- 0.7 }$ & $1681.1$ & $21.0$ & $0.13$ & $0.01$ & $7.87$ & $0.03$ & $6.05$ & $0.12$ & $0.30$ & $0.01$ & TH\\
289 & J135753.8-375905 & $1.21$ & $^{+0.11}_{- 0.14 }$ & $23.55$ & $5.36$ & $-124.7$ & $^{+3.1}_{- 3.6 }$ & $222.3$ & $^{+2.1}_{- 2.0 }$ & $52.5$ & $^{+1.4}_{- 1.5 }$ & $1253.8$ & $39.7$ & $0.35$ & $0.02$ & $7.75$ & $0.02$ & $3.74$ & $0.17$ & $0.56$ & $0.03$ & TH\\
\hline
\end{tabular}
\end{center}
\end{table}
\end{landscape}
\addtocounter{table}{-1}
\begin{landscape}
\begin{table}
\begin{center}
\caption[]{ continued. }
\label{}
\setlength{\tabcolsep}{4pt}
\tabulinesep=1.5mm
\begin{tabular}{l L{2.9cm} r@{\,}l r@{$\pm$}l r@{\,}l r@{\,}l r@{\,}l r@{$\pm$}l r@{$\pm$}l r@{$\pm$}l r@{$\pm$}l r@{$\pm$}l c}
\hline
\# & Star & \multicolumn{2}{c}{Distance}& \multicolumn{2}{c}{$RV$} &\multicolumn{2}{c}{$U$} & \multicolumn{2}{c}{$V$} & \multicolumn{2}{c}{$W$} & \multicolumn{2}{c}{$L_z$} & \multicolumn{2}{c}{$e$} & \multicolumn{2}{c}{R$_{a}$} & \multicolumn{2}{c}{R$_p$} & \multicolumn{2}{c}{Z$_{\rm max}$} & pop \\
& & \multicolumn{2}{c}{kpc} & \multicolumn{2}{c}{\kmsec} & \multicolumn{2}{c}{\kmsec} & \multicolumn{2}{c}{\kmsec} & \multicolumn{2}{c}{\kmsec} & \multicolumn{2}{c}{kpc \kmsec} & \multicolumn{2}{c}{ }   & \multicolumn{2}{c}{kpc} & \multicolumn{2}{c}{kpc}  & \multicolumn{2}{c}{kpc} &  \\
\hline
340 & J151417.0-111412 & $1.24$ & $^{+0.06}_{- 0.07 }$ & $20.76$ & $5.49$ & $-32.6$ & $^{+2.4}_{- 2.4 }$ & $238.5$ & $^{+0.5}_{- 0.5 }$ & $12.3$ & $^{+2.0}_{- 2.0 }$ & $1816.3$ & $10.1$ & $0.11$ & $0.01$ & $8.55$ & $0.05$ & $6.79$ & $0.07$ & $0.92$ & $0.03$ & TH\\
342 & J151819.6-301440 & $0.50$ & $^{+0.01}_{- 0.01 }$ & $-2.00$ & $5.36$ & $-11.3$ & $^{+2.6}_{- 2.6 }$ & $248.5$ & $^{+1.2}_{- 1.2 }$ & $2.0$ & $^{+1.2}_{- 1.2 }$ & $2026.5$ & $9.2$ & $0.07$ & $0.01$ & $9.04$ & $0.03$ & $7.81$ & $0.05$ & $0.23$ & $0.01$ & TH\\
343 & J152124.8-372354 & $1.07$ & $^{+0.05}_{- 0.05 }$ & $-41.85$ & $6.69$ & $-16.6$ & $^{+3.4}_{- 3.5 }$ & $244.8$ & $^{+2.0}_{- 2.1 }$ & $0.4$ & $^{+1.2}_{- 1.2 }$ & $1692.6$ & $17.3$ & $0.13$ & $0.01$ & $7.90$ & $0.08$ & $6.13$ & $0.08$ & $0.38$ & $0.02$ & TH\\
344 & J152214.4-373311 & $1.17$ & $^{+0.07}_{- 0.07 }$ & $16.51$ & $5.43$ & $-79.6$ & $^{+2.9}_{- 2.9 }$ & $226.6$ & $^{+1.7}_{- 1.6 }$ & $45.7$ & $^{+1.1}_{- 1.1 }$ & $1477.6$ & $20.7$ & $0.19$ & $0.01$ & $7.44$ & $0.03$ & $5.04$ & $0.11$ & $0.55$ & $0.02$ & TH\\
\hline
\end{tabular}
\end{center}
\end{table}
\end{landscape}
\addtocounter{table}{-1}
\begin{landscape}
\begin{table}
\begin{center}
\caption[]{ continued. }
\label{}
\setlength{\tabcolsep}{4pt}
\tabulinesep=1.5mm
\begin{tabular}{l L{2.9cm} r@{\,}l r@{$\pm$}l r@{\,}l r@{\,}l r@{\,}l r@{$\pm$}l r@{$\pm$}l r@{$\pm$}l r@{$\pm$}l r@{$\pm$}l c}
\hline
\# & Star & \multicolumn{2}{c}{Distance}& \multicolumn{2}{c}{$RV$} &\multicolumn{2}{c}{$U$} & \multicolumn{2}{c}{$V$} & \multicolumn{2}{c}{$W$} & \multicolumn{2}{c}{$L_z$} & \multicolumn{2}{c}{$e$} & \multicolumn{2}{c}{R$_{a}$} & \multicolumn{2}{c}{R$_p$} & \multicolumn{2}{c}{Z$_{\rm max}$} & pop \\
& & \multicolumn{2}{c}{kpc} & \multicolumn{2}{c}{\kmsec} & \multicolumn{2}{c}{\kmsec} & \multicolumn{2}{c}{\kmsec} & \multicolumn{2}{c}{\kmsec} & \multicolumn{2}{c}{kpc \kmsec} & \multicolumn{2}{c}{ }   & \multicolumn{2}{c}{kpc} & \multicolumn{2}{c}{kpc}  & \multicolumn{2}{c}{kpc} &  \\
\hline
392 & J163621.8-080636 & $0.38$ & $^{+0.01}_{- 0.01 }$ & $-50.50$ & $5.97$ & $-2.7$ & $^{+3.1}_{- 3.1 }$ & $218.3$ & $^{+0.5}_{- 0.5 }$ & $1.0$ & $^{+1.5}_{- 1.5 }$ & $1634.0$ & $5.1$ & $0.20$ & $0.01$ & $8.27$ & $0.04$ & $5.50$ & $0.04$ & $0.22$ & $0.01$ & TH\\
393 & J163828.3-245111 & $0.50$ & $^{+0.01}_{- 0.01 }$ & $3.63$ & $5.19$ & $-10.3$ & $^{+2.7}_{- 2.7 }$ & $250.3$ & $^{+0.3}_{- 0.3 }$ & $4.3$ & $^{+0.7}_{- 0.7 }$ & $2032.0$ & $2.4$ & $0.08$ & $0.01$ & $9.14$ & $0.03$ & $7.77$ & $0.04$ & $0.17$ & $0.01$ & TH\\
394 & J163947.4-663151 & $0.62$ & $^{+0.01}_{- 0.01 }$ & $-2.00$ & $5.23$ & $-12.9$ & $^{+2.2}_{- 2.2 }$ & $236.5$ & $^{+1.9}_{- 1.9 }$ & $9.0$ & $^{+0.7}_{- 0.7 }$ & $1866.6$ & $14.9$ & $0.04$ & $0.01$ & $7.95$ & $0.02$ & $7.37$ & $0.14$ & $0.14$ & $0.01$ & TH\\
395 & J163958.4-604216 & $1.44$ & $^{+0.07}_{- 0.08 }$ & $-122.91$ & $7.29$ & $70.7$ & $^{+3.4}_{- 3.3 }$ & $260.7$ & $^{+2.9}_{- 2.7 }$ & $41.8$ & $^{+0.7}_{- 0.8 }$ & $1754.5$ & $31.5$ & $0.36$ & $0.01$ & $11.17$ & $0.33$ & $5.21$ & $0.09$ & $0.65$ & $0.03$ & TH\\
\hline
\end{tabular}
\end{center}
\end{table}
\end{landscape}
\addtocounter{table}{-1}
\begin{landscape}
\begin{table}
\begin{center}
\caption[]{ continued. }
\label{}
\setlength{\tabcolsep}{4pt}
\tabulinesep=1.5mm
\begin{tabular}{l L{2.9cm} r@{\,}l r@{$\pm$}l r@{\,}l r@{\,}l r@{\,}l r@{$\pm$}l r@{$\pm$}l r@{$\pm$}l r@{$\pm$}l r@{$\pm$}l c}
\hline
\# & Star & \multicolumn{2}{c}{Distance}& \multicolumn{2}{c}{$RV$} &\multicolumn{2}{c}{$U$} & \multicolumn{2}{c}{$V$} & \multicolumn{2}{c}{$W$} & \multicolumn{2}{c}{$L_z$} & \multicolumn{2}{c}{$e$} & \multicolumn{2}{c}{R$_{a}$} & \multicolumn{2}{c}{R$_p$} & \multicolumn{2}{c}{Z$_{\rm max}$} & pop \\
& & \multicolumn{2}{c}{kpc} & \multicolumn{2}{c}{\kmsec} & \multicolumn{2}{c}{\kmsec} & \multicolumn{2}{c}{\kmsec} & \multicolumn{2}{c}{\kmsec} & \multicolumn{2}{c}{kpc \kmsec} & \multicolumn{2}{c}{ }   & \multicolumn{2}{c}{kpc} & \multicolumn{2}{c}{kpc}  & \multicolumn{2}{c}{kpc} &  \\
\hline
447 & J174342.8-161446 & $1.24$ & $^{+0.03}_{- 0.03 }$ & $-34.81$ & $6.45$ & $37.9$ & $^{+3.7}_{- 3.7 }$ & $225.2$ & $^{+0.8}_{- 0.9 }$ & $-20.4$ & $^{+0.5}_{- 0.5 }$ & $1689.4$ & $7.4$ & $0.09$ & $0.01$ & $7.53$ & $0.07$ & $6.35$ & $0.09$ & $0.30$ & $0.01$ & TH\\
448 & J174407.6-523150 & $1.95$ & $^{+0.12}_{- 0.13 }$ & $-3.41$ & $5.18$ & $5.1$ & $^{+2.5}_{- 2.6 }$ & $211.4$ & $^{+1.5}_{- 1.4 }$ & $-35.5$ & $^{+1.0}_{- 1.1 }$ & $1452.1$ & $22.5$ & $0.11$ & $0.01$ & $6.67$ & $0.06$ & $5.36$ & $0.11$ & $0.60$ & $0.03$ & TH\\
450 & J174639.8-614215 & $5.44$ & $^{+1.06}_{- 1.55 }$ & $-71.85$ & $6.59$ & $108.2$ & $^{+3.4}_{- 6.1 }$ & $153.4$ & $^{+4.3}_{- 5.2 }$ & $-37.8$ & $^{+4.0}_{- 5.3 }$ & $712.5$ & $61.4$ & $0.50$ & $0.03$ & $6.49$ & $0.14$ & $2.16$ & $0.16$ & $1.89$ & $0.42$ & TK\\
451 & J174925.5+050048 & $1.91$ & $^{+0.11}_{- 0.13 }$ & $-67.41$ & $6.88$ & $-89.3$ & $^{+4.0}_{- 4.0 }$ & $141.2$ & $^{+2.6}_{- 2.7 }$ & $37.1$ & $^{+1.5}_{- 1.4 }$ & $614.3$ & $35.2$ & $0.64$ & $0.02$ & $6.91$ & $0.04$ & $1.53$ & $0.10$ & $0.62$ & $0.04$ & TK\\
\hline
\end{tabular}
\end{center}
\end{table}
\end{landscape}
\addtocounter{table}{-1}
\begin{landscape}
\begin{table}
\begin{center}
\caption[]{ continued. }
\label{}
\setlength{\tabcolsep}{4pt}
\tabulinesep=1.5mm
\begin{tabular}{l L{2.9cm} r@{\,}l r@{$\pm$}l r@{\,}l r@{\,}l r@{\,}l r@{$\pm$}l r@{$\pm$}l r@{$\pm$}l r@{$\pm$}l r@{$\pm$}l c}
\hline
\# & Star & \multicolumn{2}{c}{Distance}& \multicolumn{2}{c}{$RV$} &\multicolumn{2}{c}{$U$} & \multicolumn{2}{c}{$V$} & \multicolumn{2}{c}{$W$} & \multicolumn{2}{c}{$L_z$} & \multicolumn{2}{c}{$e$} & \multicolumn{2}{c}{R$_{a}$} & \multicolumn{2}{c}{R$_p$} & \multicolumn{2}{c}{Z$_{\rm max}$} & pop \\
& & \multicolumn{2}{c}{kpc} & \multicolumn{2}{c}{\kmsec} & \multicolumn{2}{c}{\kmsec} & \multicolumn{2}{c}{\kmsec} & \multicolumn{2}{c}{\kmsec} & \multicolumn{2}{c}{kpc \kmsec} & \multicolumn{2}{c}{ }   & \multicolumn{2}{c}{kpc} & \multicolumn{2}{c}{kpc}  & \multicolumn{2}{c}{kpc} &  \\
\hline
510 & J184513.1-330538 & $2.66$ & $^{+0.31}_{- 0.41 }$ & $-76.12$ & $6.70$ & $51.4$ & $^{+3.7}_{- 3.7 }$ & $279.8$ & $^{+1.2}_{- 1.5 }$ & $61.9$ & $^{+1.6}_{- 1.8 }$ & $1613.5$ & $45.7$ & $0.23$ & $0.01$ & $8.84$ & $0.19$ & $5.49$ & $0.17$ & $1.40$ & $0.10$ & TK\\
512 & J184617.8-534114 & $2.18$ & $^{+0.19}_{- 0.22 }$ & $-53.41$ & $5.42$ & $18.9$ & $^{+2.8}_{- 2.9 }$ & $275.4$ & $^{+1.2}_{- 1.2 }$ & $13.2$ & $^{+1.2}_{- 1.2 }$ & $1824.3$ & $29.2$ & $0.17$ & $0.01$ & $9.18$ & $0.17$ & $6.51$ & $0.10$ & $1.03$ & $0.06$ & TH\\
513 & J184737.2-295921 & $1.18$ & $^{+0.06}_{- 0.06 }$ & $24.98$ & $5.62$ & $-34.0$ & $^{+3.3}_{- 3.3 }$ & $227.2$ & $^{+0.5}_{- 0.5 }$ & $-3.5$ & $^{+0.8}_{- 0.8 }$ & $1663.2$ & $12.6$ & $0.12$ & $0.01$ & $7.67$ & $0.08$ & $6.03$ & $0.06$ & $0.26$ & $0.01$ & TH\\
514 & J184922.2-534412 & $3.26$ & $^{+0.48}_{- 0.67 }$ & $-2.01$ & $5.47$ & $-34.8$ & $^{+2.9}_{- 3.4 }$ & $206.1$ & $^{+2.4}_{- 3.2 }$ & $12.2$ & $^{+1.1}_{- 1.1 }$ & $1156.5$ & $79.6$ & $0.16$ & $0.02$ & $5.89$ & $0.22$ & $4.26$ & $0.30$ & $1.19$ & $0.12$ & TH\\
\hline
\end{tabular}
\end{center}
\end{table}
\end{landscape}
\addtocounter{table}{-1}
\begin{landscape}
\begin{table}
\begin{center}
\caption[]{ continued. }
\label{}
\setlength{\tabcolsep}{4pt}
\tabulinesep=1.5mm
\begin{tabular}{l L{2.9cm} r@{\,}l r@{$\pm$}l r@{\,}l r@{\,}l r@{\,}l r@{$\pm$}l r@{$\pm$}l r@{$\pm$}l r@{$\pm$}l r@{$\pm$}l c}
\hline
\# & Star & \multicolumn{2}{c}{Distance}& \multicolumn{2}{c}{$RV$} &\multicolumn{2}{c}{$U$} & \multicolumn{2}{c}{$V$} & \multicolumn{2}{c}{$W$} & \multicolumn{2}{c}{$L_z$} & \multicolumn{2}{c}{$e$} & \multicolumn{2}{c}{R$_{a}$} & \multicolumn{2}{c}{R$_p$} & \multicolumn{2}{c}{Z$_{\rm max}$} & pop \\
& & \multicolumn{2}{c}{kpc} & \multicolumn{2}{c}{\kmsec} & \multicolumn{2}{c}{\kmsec} & \multicolumn{2}{c}{\kmsec} & \multicolumn{2}{c}{\kmsec} & \multicolumn{2}{c}{kpc \kmsec} & \multicolumn{2}{c}{ }   & \multicolumn{2}{c}{kpc} & \multicolumn{2}{c}{kpc}  & \multicolumn{2}{c}{kpc} &  \\
\hline
562 & J193010.8-373933 & $0.64$ & $^{+0.01}_{- 0.01 }$ & $-31.98$ & $6.58$ & $21.8$ & $^{+3.7}_{- 3.7 }$ & $249.2$ & $^{+0.1}_{- 0.1 }$ & $19.1$ & $^{+1.6}_{- 1.6 }$ & $1977.1$ & $1.8$ & $0.08$ & $0.01$ & $8.84$ & $0.08$ & $7.60$ & $0.06$ & $0.38$ & $0.02$ & TH\\
564 & J193142.8-281535 & $1.16$ & $^{+0.05}_{- 0.06 }$ & $-66.01$ & $5.86$ & $110.7$ & $^{+3.3}_{- 3.1 }$ & $262.0$ & $^{+0.7}_{- 0.8 }$ & $-20.8$ & $^{+1.3}_{- 1.5 }$ & $2056.9$ & $5.2$ & $0.30$ & $0.01$ & $12.15$ & $0.11$ & $6.51$ & $0.06$ & $0.64$ & $0.03$ & TH\\
566 & J193249.4-172704 & $1.28$ & $^{+0.07}_{- 0.08 }$ & $-117.41$ & $10.29$ & $79.8$ & $^{+5.2}_{- 5.2 }$ & $196.3$ & $^{+2.5}_{- 2.5 }$ & $68.3$ & $^{+1.8}_{- 1.8 }$ & $1394.9$ & $22.6$ & $0.32$ & $0.02$ & $8.60$ & $0.12$ & $4.39$ & $0.11$ & $1.43$ & $0.08$ & TK\\
568 & J193408.7-143216 & $1.95$ & $^{+0.15}_{- 0.18 }$ & $4.10$ & $5.27$ & $-92.2$ & $^{+3.1}_{- 3.6 }$ & $142.0$ & $^{+2.7}_{- 3.2 }$ & $15.7$ & $^{+0.9}_{- 0.9 }$ & $882.8$ & $53.9$ & $0.50$ & $0.02$ & $6.93$ & $0.07$ & $2.28$ & $0.17$ & $0.54$ & $0.03$ & TK\\
\hline
\end{tabular}
\end{center}
\end{table}
\end{landscape}
\addtocounter{table}{-1}
\begin{landscape}
\begin{table}
\begin{center}
\caption[]{ continued. }
\label{}
\setlength{\tabcolsep}{4pt}
\tabulinesep=1.5mm
\begin{tabular}{l L{2.9cm} r@{\,}l r@{$\pm$}l r@{\,}l r@{\,}l r@{\,}l r@{$\pm$}l r@{$\pm$}l r@{$\pm$}l r@{$\pm$}l r@{$\pm$}l c}
\hline
\# & Star & \multicolumn{2}{c}{Distance}& \multicolumn{2}{c}{$RV$} &\multicolumn{2}{c}{$U$} & \multicolumn{2}{c}{$V$} & \multicolumn{2}{c}{$W$} & \multicolumn{2}{c}{$L_z$} & \multicolumn{2}{c}{$e$} & \multicolumn{2}{c}{R$_{a}$} & \multicolumn{2}{c}{R$_p$} & \multicolumn{2}{c}{Z$_{\rm max}$} & pop \\
& & \multicolumn{2}{c}{kpc} & \multicolumn{2}{c}{\kmsec} & \multicolumn{2}{c}{\kmsec} & \multicolumn{2}{c}{\kmsec} & \multicolumn{2}{c}{\kmsec} & \multicolumn{2}{c}{kpc \kmsec} & \multicolumn{2}{c}{ }   & \multicolumn{2}{c}{kpc} & \multicolumn{2}{c}{kpc}  & \multicolumn{2}{c}{kpc} &  \\
\hline
620 & J204812.3-281705 & $4.27$ & $^{+0.78}_{- 1.15 }$ & $-70.30$ & $6.14$ & $126.3$ & $^{+5.5}_{- 7.1 }$ & $174.5$ & $^{+5.1}_{- 5.7 }$ & $-8.6$ & $^{+3.6}_{- 4.6 }$ & $974.1$ & $120.6$ & $0.36$ & $0.04$ & $7.45$ & $0.09$ & $3.46$ & $0.35$ & $3.26$ & $0.51$ & TK\\
621 & J204942.0+074816 & $1.32$ & $^{+0.07}_{- 0.08 }$ & $4.96$ & $5.53$ & $6.9$ & $^{+1.4}_{- 1.4 }$ & $245.3$ & $^{+2.5}_{- 2.5 }$ & $2.0$ & $^{+1.2}_{- 1.2 }$ & $1925.9$ & $20.3$ & $0.03$ & $0.01$ & $8.25$ & $0.15$ & $7.72$ & $0.04$ & $0.49$ & $0.02$ & TH\\
622 & J204954.3-693630 & $1.44$ & $^{+0.07}_{- 0.07 }$ & $88.93$ & $7.72$ & $-6.5$ & $^{+2.9}_{- 3.1 }$ & $131.3$ & $^{+2.6}_{- 2.4 }$ & $-78.3$ & $^{+2.6}_{- 2.6 }$ & $923.5$ & $28.6$ & $0.46$ & $0.01$ & $7.65$ & $0.01$ & $2.82$ & $0.08$ & $2.30$ & $0.16$ & TK\\
624 & J205153.5-550734 & $2.96$ & $^{+0.33}_{- 0.43 }$ & $-4.83$ & $5.39$ & $37.3$ & $^{+2.6}_{- 2.7 }$ & $219.9$ & $^{+1.2}_{- 1.5 }$ & $-62.1$ & $^{+2.8}_{- 3.7 }$ & $1368.4$ & $44.7$ & $0.20$ & $0.02$ & $7.71$ & $0.07$ & $5.16$ & $0.17$ & $2.39$ & $0.21$ & TK\\
\hline
\end{tabular}
\end{center}
\end{table}
\end{landscape}
\addtocounter{table}{-1}
\begin{landscape}
\begin{table}
\begin{center}
\caption[]{ continued. }
\label{}
\setlength{\tabcolsep}{4pt}
\tabulinesep=1.5mm
\begin{tabular}{l L{2.9cm} r@{\,}l r@{$\pm$}l r@{\,}l r@{\,}l r@{\,}l r@{$\pm$}l r@{$\pm$}l r@{$\pm$}l r@{$\pm$}l r@{$\pm$}l c}
\hline
\# & Star & \multicolumn{2}{c}{Distance}& \multicolumn{2}{c}{$RV$} &\multicolumn{2}{c}{$U$} & \multicolumn{2}{c}{$V$} & \multicolumn{2}{c}{$W$} & \multicolumn{2}{c}{$L_z$} & \multicolumn{2}{c}{$e$} & \multicolumn{2}{c}{R$_{a}$} & \multicolumn{2}{c}{R$_p$} & \multicolumn{2}{c}{Z$_{\rm max}$} & pop \\
& & \multicolumn{2}{c}{kpc} & \multicolumn{2}{c}{\kmsec} & \multicolumn{2}{c}{\kmsec} & \multicolumn{2}{c}{\kmsec} & \multicolumn{2}{c}{\kmsec} & \multicolumn{2}{c}{kpc \kmsec} & \multicolumn{2}{c}{ }   & \multicolumn{2}{c}{kpc} & \multicolumn{2}{c}{kpc}  & \multicolumn{2}{c}{kpc} &  \\
\hline
677 & J230724.0-605823 & $4.97$ & $^{+0.85}_{- 1.21 }$ & $173.44$ & $18.02$ & $-212.0$ & $^{+7.2}_{- 9.9 }$ & $146.8$ & $^{+7.0}_{- 7.1 }$ & $-87.1$ & $^{+7.4}_{- 7.4 }$ & $1019.4$ & $110.1$ & $0.65$ & $0.03$ & $21.65$ & $2.65$ & $4.61$ & $0.10$ & $17.45$ & $3.60$ & H\\
678 & J230728.9-602936 & $1.81$ & $^{+0.14}_{- 0.16 }$ & $12.29$ & $5.51$ & $-25.9$ & $^{+1.4}_{- 1.4 }$ & $196.0$ & $^{+1.5}_{- 1.9 }$ & $14.7$ & $^{+2.5}_{- 2.4 }$ & $1461.3$ & $26.7$ & $0.20$ & $0.01$ & $7.72$ & $0.03$ & $5.16$ & $0.12$ & $1.44$ & $0.07$ & TK\\
679 & J230808.7-600954\,$\dagger$ & $4.45$ & $^{+0.53}_{- 0.69 }$ & $184.40$ & $10.75$ & $-46.5$ & $^{+4.3}_{- 5.1 }$ & $-1.4$ & $^{+6.2}_{- 7.7 }$ & $-104.3$ & $^{+5.3}_{- 5.5 }$ & $-106.6$ & $83.0$ & $0.88$ & $0.04$ & $7.98$ & $0.24$ & $0.52$ & $0.19$ & $6.81$ & $0.53$ & H\\
680 & J230830.4-113847 & $1.43$ & $^{+0.10}_{- 0.12 }$ & $-43.31$ & $5.79$ & $-56.2$ & $^{+1.5}_{- 1.6 }$ & $197.5$ & $^{+1.6}_{- 1.7 }$ & $46.8$ & $^{+3.1}_{- 3.1 }$ & $1623.9$ & $17.9$ & $0.25$ & $0.01$ & $9.12$ & $0.06$ & $5.53$ & $0.09$ & $1.66$ & $0.07$ & TK\\
\hline
\end{tabular}
\end{center}
\end{table}
\end{landscape}


\label{lastpage}
\end{document}